\documentclass[twocolumn,10pt,aps,prd,floatfix,nofootinbib]{revtex4-1}

\newcommand{\bk}[0]{\mathbf{k}}
\newcommand{\bP}[0]{\mathbf{P}}

\usepackage{amsmath}
\usepackage{amssymb}
\usepackage{amsfonts}
\usepackage{graphicx}
\usepackage[colorlinks=true,linkcolor=blue,citecolor=blue,urlcolor=blue]{hyperref}
\usepackage{array}
\usepackage{booktabs}
\usepackage{psfrag}
\usepackage{csquotes}

\providecommand{\abs}[1]{\left|#1\right|}
\providecommand{\ep}[1]{{e}^{#1}}

\newcommand{\eq}[1]{Eq.~\eqref{#1}}
\newcommand{\om}[0]{\omega}

\begin{document}
\title{Dissipative fields in de Sitter and black hole spacetimes: 
\texorpdfstring{\\}{}
Quantum entanglement due to pair production and dissipation}
\author{Julian Adamek}
\email{julian.adamek@unige.ch}
\affiliation{D\'epartement de Physique Th\'eorique \& Center for Astroparticle Physics, 
         \\Universit\'e de Gen\`{e}ve, 24 Quai Ernest Ansermet, 1211 Gen\`{e}ve 4, Switzerland}
\author{Xavier Busch}
\email[]{xavier.busch@th.u-psud.fr}
\author{Renaud Parentani}
\email[]{renaud.parentani@th.u-psud.fr}
\affiliation{Laboratoire de Physique Th\'eorique, CNRS UMR 8627, B{\^{a}}t.\ 210,
         \\Universit\'e Paris-Sud 11, 91405 Orsay CEDEX, France}

\begin{abstract}
For free fields, pair creation in expanding universes is associated with the building up of correlations that lead to nonseparable states, i.e., quantum mechanically entangled ones. For dissipative fields, i.e., fields coupled to an environment, there is a competition between the squeezing of the state and the coupling to the external bath. We compute the final coherence level for dissipative fields that propagate in a two-dimensional de Sitter space, and we characterize the domain in parameter space where the state remains nonseparable. We then apply our analysis to (analogue) Hawking radiation by exploiting the close relationship between Lorentz violating theories propagating in de Sitter and black hole metrics. We establish the robustness of the spectrum and find that the entanglement among Hawking pairs is generally much stronger than that among pairs of quanta with opposite momenta.
\end{abstract}
\maketitle

\section{Introduction}

The propagation of quantum fields in expanding cosmological backgrounds leads to the spontaneous creation of pairs of particles with opposite momenta~\cite{birrell1984quantum}. For free fields, relativistic or dispersive, this pair creation (also called the dynamical Casimir effect in condensed matter physics, see e.g., Refs.~\cite{Carusotto:2009re,PhysRevLett.109.220401}) is associated with the building up of nonlocal correlations that lead to quantum mechanically entangled states~\cite{Campo:2003pa,Campo:2005sv}. To define these states without ambiguity, we shall use the notion of nonseparability~\cite{Werner:1989}, see Appendix~\ref{app:CSclass}. For dissipative fields, i.e.,\ fields coupled to an environment, there is a competition between the squeezing of the state, which increases the strength of the correlations, and the coupling to the external bath, which reduces it~\cite{Campo:2005sy,Prokopec:2006fc,Campo:2008ju,*Campo:2008ij}. 

Our principal aim is to study this competition. We shall work both in time-dependent (cosmological) settings and with stationary metrics. For simplicity and definiteness, we consider fields that propagate in a two-dimensional de~Sitter space and display dissipative effects above a certain momentum threshold $\Lambda$. For these fields, the final coherence level is constant and well defined. We characterize the domain in parameter space where the final state is nonseparable. The parameters are the mass of the field, the temperature of the environment, and the ratio $\Lambda / H$, where $H$ is the Hubble constant. Since the dissipative/dispersive effects we are considering are suppressed in the infrared, our models can be conceived as providing a phenomenological approach to theories of quantum gravity, such as Ho\v{r}ava-Lifshitz gravity~\cite{Horava:2009uw}, where Lorentz invariance is violated at high energy. In these theories, dissipative effects will necessarily appear through radiative corrections~\cite{Jacobson:2005bg}. We also recall that in condensed matter, the spectrum of quasiparticles often displays dissipation above a certain threshold. Hence, our model can also be viewed as a toolbox to compute the consequences of dissipation on pair production and parametric amplification found, e.g., in the superfluid of polaritons studied in Ref.~\cite{Gerace:2012an}.

The interest in working in de Sitter space is twofold. On the one hand, the analysis of the state can be done in terms of homogeneous modes and pair creation of quanta with opposite momentum. On the other hand, the state can also be analyzed in terms of stationary modes and thermal-like effects associated with the Gibbons-Hawking temperature~\cite{birrell1984quantum}. It is rather clear that the homogeneous representation in de Sitter can be conceived as an approximation to e.g.,\ slow roll inflation, see Refs.~\cite{Adamek:2008mp,Macher:2008yq}. What is less obvious is that de~Sitter also provides a reliable approximation to describe dissipative fields propagating in black hole metrics. Indeed, when the ultraviolet scale $\Lambda$ is well separated from the surface gravity of the black hole, the dissipative aspects of typical Hawking quanta all occur in the near horizon region, which can be mapped into a portion of de~Sitter space (when the Hubble constant is matched to the surface gravity). As a result, the state evaluated in a black hole metric can be well approximated by the corresponding one evaluated in de~Sitter. In this respect, the present paper follows up on our former work~\cite{Busch:2012ne} where we studied this correspondence for dispersive fields. The reader unfamiliar with field propagation in de Sitter space will find in that work all necessary information. 

This paper is organized as follows. In Sec.~\ref{sec:actionsetting} we present the action which engenders dissipative effects, and we discuss the residual symmetries found in de Sitter space when considering such theories. In Sec.~\ref{sec:homogen}, exploiting the homogeneity of de Sitter, we compute the spectral properties and the correlations of pairs with opposite momenta. In Sec.~\ref{sec:statio}, exploiting the stationarity, we compute the deviations with respect to the Gibbons-Hawking temperature. We apply our model to black holes in Sec.~\ref{sec:BHdScorr}, and we conclude in Sec.~\ref{sec:conclusions}. We work in units where $\hbar =c=1$. 

\section{Dissipative and dispersive fields}
\label{sec:actionsetting}

\subsection{Covariant settings}

We study a scalar field $\phi$ that has a standard relativistic behavior at low energy but displays dispersion and dissipation at high energy, thereby violating (local) Lorentz invariance. While high-energy dispersion is rather easily introduced and has been studied in many papers both in cosmological settings~\cite{Martin:2000xs,Niemeyer:2000eh,Niemeyer:2001qe,Macher:2008yq} and black hole metrics~\cite{Unruh:1994je,Brout:1995wp}, see e.g., Ref.~\cite{Jacobson:1999zk} for a review, dissipation has received comparatively much less attention. When preserving unitarity and general covariance, dissipation is also technically more difficult to handle. To do so in simple terms, following~\cite{Parentani:2007uq}, we introduce dissipation by coupling $\phi$ to some environmental degrees of freedom $\psi$, and the action of the entire system $S_{\rm tot} = S_\phi + S_\psi + S_{\rm int}$ is taken quadratic in $\phi,\psi$, as in models of atomic radiation damping~\cite{Aichelburg1976264} and quantum Brownian motion~\cite{Unruh:1989dd}. Again for reasons of simplicity, we shall work in $1+1$ dimensions. The reader interested in four-dimensional models may consult~\cite{Adamek:2008mp}, where there is a phenomenological study of inflationary spectra in dissipative models. 

In the present work, we consider dispersion relations that contain both dispersive and dissipative effects. These relations can be parametrized by two real functions $\Gamma,f$ as
\begin{equation}
\label{eq:dispersion}
 \Omega^2 + 2 i \Gamma \Omega= m^2 + P^2+ f = F^2~, 
\end{equation}
where $\Gamma(P^2)> 0$ is the damping rate, and $f(P^2) $ describes dispersive effects. To recover a relativistic behavior in the infrared, a typical behavior would be $\Gamma \sim P^2$ and $f \sim P^4$ for $P^2 \to 0$. In \eq{eq:dispersion}, $\Omega$ and $P^2$ are, respectively, the proper frequency and the proper momentum squared as measured in the ``preferred'' frame~\cite{Corley:1996ar}, i.e., the frame used to implement the dispersion relation. In condensed matter systems, it is provided by the medium. Instead, in the phenomenological approach to Lorentz violating effects we are pursuing, it should be given from the outset, either as a dynamical field endowed with an action~\cite{Eling:2004dk,Horava:2009uw}, or as a background field (as we shall do). To describe it in covariant terms, following Ref.~\cite{Jacobson:1996zs}, we introduce both the unit timelike vector field $u$ which describes the flow of preferred observers, and the unit spacelike vector field $s$ which is orthogonal to $u$. In terms of these, one has $\Omega = u^\mu p_\mu$ and $P^2 = (s^\mu p_\mu)^2$ where $p_\mu$ is the momentum of the particle in an arbitrary coordinate system. In two dimensions, the metric $\mathsf{g}_{\mu \nu}$ can be written as $ \mathsf{g}_{\mu \nu} = -u_\mu u_\nu + s_\mu s_\nu$ which expresses that $u$ and $s$ are orthonormal vectors.

We now consider a unitary model which implements \eq{eq:dispersion}. This model is not unique but can be considered as the simplest one, as shall be made clear below. In covariant terms, the total action $S_{\rm tot} = S_\phi + S_\psi + S_{\rm int}$ is
\begin{equation}
\begin{split}
\label{eq:covaction}
S_{\rm tot} = &\frac{1}{2} \int\! d^2 \left[-\mathsf{g}^{\mu\nu} \nabla_\mu \phi \nabla_\nu \phi \! -\! m^2 \phi^2 \! -\! \phi f\! \left(-{ \nabla}_s^2 \right) \phi \right]\\
&+ \frac{1}{2} \int\! d^2 \int\! dq \left[\left( \nabla_u \psi_q\right)^2 - \left(\pi \Lambda q\right)^2 \psi_q^2\right] \\
& + \int\! d^2 \left[\bigg( \gamma \left( \nabla_s \right) \phi\bigg) \left(  \nabla_u \int\!\! dq \,  \psi_q \right)\right]~,
\end{split}
\end{equation}
where $d^2 = d^2 \mathsf x \sqrt{ - \mathsf g(\mathsf x)}$ is the covariant measure. In the first line, $S_\phi$ is the standard action of a massive scalar field, apart from the last term which introduces the high frequency dispersion described by $f(P^2)$. In two dimensions, the self-adjoint operator which implements $P^2$ is $-\nabla_s^2 \doteq  \nabla_s^\dagger  \nabla_s$, where $ \nabla_s = s^\mu \nabla_\mu $ is an anti-self-adjoint operator (when $u$ is a freely falling frame), $ \nabla_s^\dagger = - \nabla_\mu s^\mu$ its adjoint, and $ \nabla_\mu $ the covariant derivative. A four-dimensional version of this model can be found in~\cite{Parentani:2007uq}.

The second action, that of the $\psi$ field, contains the extra dimensionless parameter $q$, which can be considered as a wave number in some extra dimension. Its role is to guarantee that the environment degrees of freedom are dense, something necessary to engender dissipative effects when coupling $\psi$ to $\phi$~\cite{Parentani:2007uq,Unruh:1989dd}. The role of the frequency $\Lambda$ is to set the ultraviolet scale where dissipative effects become important. The kinetic term of $\psi$ is governed by the anti-self adjoint operator $ \nabla_u \doteq -(u^\mu \nabla_\mu + \nabla_\mu u^\mu) / 2$ which implements $ \Omega = u^\mu p_\mu$. We notice that there is no spatial derivative acting on $\psi$. This means that the quanta of $\psi$ are at rest in the preferred frame. This restriction can easily be removed by adding the term $c^2_\psi ( \nabla_s \psi)^2$ which associates to $c_\psi$ the group velocity of the low $q$ quanta. Including this term leads to much more complicated equations because dissipative effects are then described by a nonlocal kernel, as shall be briefly discussed after \eq{eq:Gammaasgamma}. For reasons of simplicity, we shall work with  $c_\psi = 0$ which gives a local kernel. Moreover, in homogeneous universes $c_\psi = 0$ also implies that the $\Psi$-modes are not parametrically amplified by the cosmological expansion. When working with given functions $\Gamma(P^2)$ and $f(P^2)$, we do not expect that the complications associated with $c_\psi \neq 0$ will qualitatively modify the effective behavior of $\phi$, at least when $\Lambda$ is well separated from the Hubble scale. 

The interaction between the two fields is given by the third action. The strength and the momentum dependence of the coupling is governed by the function $\gamma(P)$ which has the dimension of a momentum. Its role is to engender the decay rate $\Gamma$ entering \eq{eq:dispersion}. The last two actions possess peculiar properties which have been adopted to obtain simple equations of motion. These are
\begin{subequations}
\label{eq:eomcov}
\begin{align}
\label{eq:eomcovphi}
\left [\nabla_\mu u^\mu u^\nu \nabla_\nu +F^2(- \nabla_s^2) \right ] \phi & = \gamma( \nabla_s^\dagger)  \nabla_u  \int dq \psi_q~, \\
\label{eq:eomcovpsi}
\left [ \nabla_u^2+(\pi \Lambda q)^2 \right ] \psi_q &= -  \nabla_u \gamma( \nabla_s )  \phi~.
\end{align}
\end{subequations}
The solution to the second equation is 
\begin{equation}
\psi_q(\mathsf x') = \psi_q^0(\mathsf x') - \int d^2 \, G_q(\mathsf x',\mathsf x) \nabla_{u} \gamma(\nabla_s )  \phi(\mathsf x)~,
\label{psiq}
\end{equation}
where $\psi_q^0$ is a homogeneous solution, and where the driven solution is governed by $G_q(\mathsf x,\mathsf x')$, the retarded Green function of $\psi_q$. When injecting $\psi_q$ in the rhs of the first equation, one obtains the equation of $\phi$ driven by $\psi_q^0$. The general solution can be written as $\phi = \phi^{\rm dec} + \phi^{\rm dr}$, where the decaying part is a homogeneous solution, and where the driven part is given by
\begin{equation}
\begin{split}
\phi^{\rm dr}(\mathsf x')  =\!\! \int d^2 \, G_{\rm ret}(\mathsf x',\mathsf x)  \gamma(\nabla^\dagger_s) \nabla_u  \int dq \psi_q^0(\mathsf x)~. \\ 
\end{split}
\label{eq:phidr}
\end{equation}
In a general Gaussian $\phi-\psi$ model, the retarded Green function $G_{\rm ret}$ would obey a nonlocal equation, i.e., an integro-differential equation. We have adjusted the properties of $S_\psi$ and $S_{\rm int}$ precisely to avoid this. Two properties are essential. Firstly, at fixed $q$ and along the orbits of $u$, Eq.~\eqref{eq:eomcovpsi} reduces to that of a driven harmonic oscillator. This can be seen by introducing the coordinates $(\tau,z)$ defined by $u^\mu \partial_\mu = -\partial_{\tau\vert z}$ where $z$ is a spatial coordinate which labels the orbits of $u$. Then, $ \nabla_u$ applied on scalars is
\begin{equation}
\label{eq:Duincov}
 \nabla_u   =a^{-1/2} \, \partial_{\tau\vert z} \,  a^{1/2} ~,
\end{equation} 
where $a(\tau,z) \doteq \ep{ \int^\tau d\tau' \Theta(\tau',z)}$, and where $\Theta \doteq - \nabla_\mu u^\mu$ is the expansion of $u$. Hence the rescaled field 
\begin{equation}
\label{eq:psirenorm}
\Psi_q(\tau,z) \doteq  \sqrt{a(\tau,z)} \, \psi_q(\tau,z)
\end{equation} 
obeys the equation of an oscillator of constant frequency $\pi \Lambda\abs{q}$. Secondly, when summed over $q$, the retarded Green function of $\psi$ obeys~\cite{Parentani:2007uq}
\begin{equation}
 \nabla_u \int_{-\infty}^{\infty} dq \, G_q(\mathsf x,\mathsf x') = \frac{\delta^2(\mathsf x-\mathsf x')}{\Lambda} ~,
\label{eq:loc}
\end{equation}
where $\delta^2(\mathsf x-\mathsf x')$ is the covariant Dirac delta, i.e., $\int d^2 f(\mathsf x) \delta^2(\mathsf x-\mathsf x') = f(\mathsf x') $. \eq{eq:loc} guarantees that the differential operator encoding dissipation is local. Namely, when inserting $\psi_q$ of \eq{psiq} in Eq.~\eqref{eq:eomcovphi}, one finds
\begin{equation}
\label{eq:eomcovariant}
\begin{split}
\Box_{\rm diss}&\, \phi  = \gamma(  \nabla_s^\dagger)   \nabla_u  \int dq \psi_q^0 ~, \\
\end{split}
\end{equation}
with the local differential operator 
\begin{equation}
\label{eq:boxdiss}
\begin{split}
\Box_{\rm diss}&\! \doteq \! \left [\nabla_\mu u^\mu u^\nu \nabla_\nu \! + F^2(- \nabla_s^2) + \frac{\gamma(  \nabla_s^\dagger)  \nabla_{u} \gamma(  \nabla_s) }{\Lambda}  \right ]\! ~.
\end{split}
\end{equation}

One can now verify that the WKB solutions of $\Box_{\rm diss}\, \phi = 0$ are governed by a Hamilton-Jacobi action which obeys the dispersion relation of Eq.~\eqref{eq:dispersion} with 
\begin{equation}
\label{eq:Gammaasgamma}
\Gamma =  \abs{\gamma}^2 / 2\Lambda~, 
\end{equation}
see Appendix~\ref{app:BHdS} for more details. The reader can also verify that any modification of the actions $S_\psi$ and $S_{\rm int}$ leads to the replacement of $\Box_{\rm diss}$ by a nonlocal operator. When considered in homogeneous and static situations, this is not problematic because one can work with Fourier modes in both space and time. However when considered in nonhomogeneous and/or nonstatic backgrounds, it becomes hopeless to solve such an equation by analytical methods. 

In our model, the retarded Green function thus obeys 
\begin{equation}
\begin{split}
\Box_{\rm diss} \, G_{\rm ret}(\mathsf x,\mathsf x') = \delta^2 (\mathsf x-\mathsf x') ~,
\label{locEq}
\end{split}
\end{equation}
and vanishes when $\mathsf x$ is in the past of $\mathsf x'$, where the past is defined with respect to the foliation introduced by the $u$ field. When canonically quantizing $\phi$ and $\psi$, since our action is Gaussian, the commutator $G_\mathrm{c}(\mathsf x,\mathsf x') \! \doteq \! [\hat\phi(\mathsf x),\hat\phi(\mathsf x')]$ is independent of $\hat \rho_{\rm tot}$, the state of the entire system. Moreover, it is related to $G_{\rm ret}$ in the usual way
\begin{equation}
\label{eq:GcfromGret}
\begin{split}
-i\, G_\mathrm{c}(\mathsf x,\mathsf x') = G_{\rm ret}(\mathsf x,\mathsf x') - G_{\rm ret}(\mathsf x',\mathsf x)~. 
\end{split}
\end{equation}

In this paper we only consider Gaussian states. This implies~\cite{mandel1995optical,leonhardt1997measuring} that the density matrix $\hat \rho_{\rm tot}$, and all observables, are completely determined by the anti-commutator of $\hat \phi$,
\begin{equation}
\label{eq:Gacdef}
  G_{\rm ac}(\mathsf x,\mathsf x')\!  \doteq\! {\rm Tr} \!\left (\hat \rho_{\rm tot} \, \{\hat \phi(\mathsf x),\hat \phi(\mathsf x')\}\right )~,
\end{equation}
that of $\hat \psi$, and the mixed one containing $\hat \phi$ and $\hat \psi$. Decomposing the field operator $\hat \phi = \hat \phi^{\rm dec} + \hat \phi^{\rm dr}$, $G_{\rm ac}$ splits into three terms. The first one involves only $\hat \phi^{\rm dec}$, the second contains both $\hat \phi^{\rm dec}$ and $\hat \phi^{\rm dr}$, and the last only $\hat \phi^{\rm dr}$. When assuming that the initial conditions are imposed in the remote past, because of dissipation, only the last one is relevant. Using \eq{eq:phidr}, it is given by\footnote{This equation can be viewed as a Gaussian version of the Keldysh (or Kadanoff - Baym) equation. It arises in many contexts, see for example~\cite{Hu:1999mm} in stochastic gravity, \cite{PhysRevB.50.5528} in resonant-tunneling systems, and~\cite{Bode2012} in Nanoelectromechanics.} 
\begin{equation}
\label{eq:acgrennfunction}
G_{\rm ac}^{\rm dr}(\mathsf x,\mathsf x')\!  =   \int\!\!\!\int \! \!   d^2_1 d^2_2 \, G_{\rm ret}(\mathsf x,\mathsf x_1) G_{\rm ret}(\mathsf x',\mathsf x_2) N(\mathsf x_1,\mathsf x_2)~,
\end{equation}
where the noise kernel is
\begin{equation}
\label{eq:disskernel}
\begin{split}
N(\mathsf x,\mathsf x') \doteq &   \,  \gamma(\nabla^\dagger_s)  \nabla_{u}\,  \gamma(\nabla'^\dagger_s) \nabla'_{u} \\
&   \int\!\!\!\int dq dq' {\rm Tr} \!\left (\hat \rho_{\rm tot} \{\hat \psi_q^{0}(\mathsf x),\hat \psi_{q'}^{0}(\mathsf x')\}\right )~.
\end{split}
\end{equation} 
In Secs.~\ref{sec:homogen} and \ref{sec:statio}, we compute $G_{\rm ac}^{\rm dr}$ and extract from it pair creation probabilities and Hawking--like effects taking place in de Sitter space. 

\subsection{Affine group in de Sitter space}

The two dimensional de~Sitter space possesses three Killing vector fields that generate the algebra of the Lie group $SO(1,2)$. Imposing that the action is invariant under the full group precludes ultraviolet dispersive and dissipative effects such as those of \eq{eq:dispersion}, see Appendix~A in Ref.~\cite{Busch:2012ne} for the proof. Since we want to work with \eq{eq:dispersion}, we must break (at least) one of these symmetries. As in Refs.~\cite{Eling:2006xg,Busch:2012ne}, we preserve the invariance under a two dimensional sub-group which corresponds to the affine group. Its algebra is generated by the Killing fields $K_z$ and $K_t$. Using the cosmological coordinates $t,z$ of the Poincar\'e patch
\begin{equation}
\label{eq:FRW}
 ds^2 = -dt^2 + e^{2 H t} dz^2 = \frac{1}{H^2 \eta^2} \left[-d\eta^2 + dz^2\right]~,
\end{equation}
$K_z = \partial_z$ generates translations in $z$ and expresses the homogeneity of the sections $t = \mathit{cst.}$, whereas $K_t = \partial_t -H z \partial_z$ expresses the stationarity of de Sitter. Using $X = \ep{H t} z$, this symmetry becomes manifest,
\begin{equation}
\label{eq:PG}
 ds^2 = -dt^2 + \left(dX - H X dt\right)^2~.
\end{equation}
Considering the two Killing fields $K_z$ and $K_t$, there is only one unit timelike freely falling field which commutes with both of them. We call it $u^{\mathrm{ff}}$, and we call $s_{\mathrm{ff}}$ the spatial unit orthogonal vector $u^{\mathrm{ff}} \cdot s_{\mathrm{ff}} = 0$. Then the coordinates $t,X$ are both {\it invariantly} defined in terms of $u^{\mathrm{ff}},\, s_{\mathrm{ff}}$ by $dt = u^{\mathrm{ff}}_\mu dx^\mu$, $\partial_X|_t = s_{\mathrm{ff}}^\mu \partial_\mu$.  

Imposing that the action of Eq.~\eqref{eq:eomcovariant} be invariant under the affine group requires that the preferred fields $u$ and $s$ commute with $K_t$ and $K_z$. This fixes $u$ and $s$ up to a boost, see Appendix~\ref{App:Prep}. For simplicity, in what follows, we work with $u = u^{\mathrm{ff}}$. In this case, the preferred frame coincides with the cosmological one, and the orbits of $u$ are $z = \mathit{cst}$. 

We also impose that the states $\hat \rho_{\rm tot}$ are invariant under the affine group. This is analogous to the restriction to the so-called $\alpha-$vacua which are invariant under the full de Sitter group~\cite{Schomblond:1976xc,Mottola:1984ar}. This means that $G_{\rm ac}$, $G_{\rm ret}$ and $N$ of \eq{eq:acgrennfunction} will be invariant under both $K_t$ and $K_z$. However, because the commutator $ [ K_z , K_t] = - H K_z$ does not vanish, one cannot simultaneously diagonalize $K_z$ and $K_t$. This leads to two different ways to express the two-point functions, either at fixed wave number $\bk = -i \partial_{z\vert t}$, or at fixed frequency $\omega = i\partial_{t\vert X} $. Explicitly, one has 
\begin{subequations}
\label{eq:fourierkt}
\begin{align}
\label{eq:fourierk}
 G_{\rm any}^k(\eta,\eta')&\doteq \int d\Delta z \, \ep{-i \bk \Delta z } \, G_{\rm any}(\Delta z ,\eta,\eta')~,\\
\label{eq:fouriert}
G_{\rm any}^{\omega}(X,X') &\doteq \int d\Delta t \, \ep{i \omega \Delta t } \, G_{\rm any}(\Delta t ,X,X')~, 
\end{align}
\end{subequations}
where $k = \abs{\bk} $, and where the ``any'' subscript indicates that these Fourier transforms apply to any two-point function which is invariant under the affine group. (In \eq{eq:fourierk}, $G^k$ only depends on $k$ because we impose isotropy.) 

What is specific to this group is that the two symmetries combine in a nontrivial way, and imply that two-point functions only depend on two quantities, and not three, as it is generally the case in homogeneous or stationary metrics. In the homogeneous representation, it implies that the product $ k\,  G_{\rm any}^k(\eta,\eta')$ only depends on the physical momenta 
$P = -H k \eta $, $P' = - H k\eta'$. Hence, in what follows, we work in the $P$-representation with
\begin{equation}
\label{Prep2ptfn}
G_{\rm any} (P,P') \doteq \frac{k}{H} G_{\rm any}^k(\eta,\eta') ~.
\end{equation}
To reach this representation when starting from the stationary $G_{\rm any}^{\omega}(X,X')$ is more involved, and is explained in Appendix~\ref{App:Prep}. 

\section{Homogeneous picture}
\label{sec:homogen}

\subsection{Dissipation and nonseparability} 
\label{sec:homoentangle}

In this section, we decompose the fields in Fourier modes of fixed $\bk$. This representation is suitable for studying the cosmological pair-creation effects induced by the expansion $a(t) = e^{Ht} = - 1/H\eta$. 

To express the outcome of dissipation in standard terms, we exploit the fact that Lorentz invariance is recovered in the infrared, for momenta $P = k e^{-Ht} \ll \Lambda$. In this limit, since $\Gamma$ and $f$ of \eq{eq:dispersion} are negligible, the $\bk$ components of $\hat \phi$ decouple from $\hat \psi$, and obey a relativistic wave equation. Hence, the $\bk$ component of the (driven) field operator of \eq{eq:phidr} can be decomposed in the \textit{out} basis as
\begin{equation}
\label{eq:phi_k-decomposition}
\hat \phi_\bk(t) \underset{t\to \infty}{\sim} \hat{\mathrm{a}}_\bk \varphi_k(t) + \hat{\mathrm{a}}^\dagger_{-\bk} \varphi^\ast_k(t)~,
\end{equation}
where the \textit{out} modes obey the scalar wave equation and satisfy the standard positive frequency condition at late time. This means that the (reduced) state of $\hat \phi$ (obtained by tracing over $\hat \psi$) can be asymptotically described in terms of conventional excitations with respect to the asymptotic \textit{out}-vacuum. 

The \textit{out} operators $ \hat{\mathrm{a}}_\bk, \hat{\mathrm{a}}^\dagger_{\bk}$ obey the standard commutation rule $[\hat{\mathrm{a}}_\bk, \hat{\mathrm{a}}^\dagger_{\bk'}] = \delta(\bk - \bk')$. For notational simplicity, we omit the $\delta(\bk-\bk')$ when writing two-point functions because it is common to all of them since we only consider homogeneous states. For instance, $ {\rm Tr} \!\left (\hat \rho_{\rm tot} \, \{\hat \phi_\bk^\dagger, \hat\phi_{\bk'}\}\right )=\delta(\bk-\bk') \times G_{\rm ac}^k$. Using \eq{eq:phi_k-decomposition}, the coefficient of the $\delta$ function is
\begin{multline}
\label{eq:phi-Ga} 
\left.G_{\rm ac}^k(t,t)\! \right|_{t \rightarrow \infty}= 2 \left[2 n_{k} + 1\right] \vert  \varphi_k(t) \vert^2 + 4  \mathrm{Re} (c_{k} \varphi_k^2(t) )~, 
\end{multline}
 where 
\begin{subequations}
\label{eq:nkckdef}
\begin{align}
 n_{k} &\doteq {\rm Tr} \!\left (\hat \rho_{\rm tot} \, \hat{\mathrm{a}}^\dagger_\bk \hat{\mathrm{a}}_{\bk}\right )  = {\rm Tr} \!\left (\hat \rho_{\rm tot} \, \hat{\mathrm{a}}^\dagger_{-\bk} \hat{\mathrm{a}}_{-\bk}\right ) ~,\\
 c_{k} &\doteq  {\rm Tr} \!\left (\hat \rho_{\rm tot} \, \hat{\mathrm{a}}_{-\bk} \hat{\mathrm{a}}_{\bk}\right )~.
 \end{align}
\end{subequations}
The mean number of asymptotic outgoing particles is $n_k > 0$, whereas the complex number $c_k$ characterizes the strength of the correlations between particles of opposite wavenumber. The relative magnitude of this number leads to the notion of nonseparability. 

To explain this, we recall that the correlations weighted by $c_k$ obey the following Cauchy-Schwartz inequality, 
\begin{eqnarray}
\vert c_k \vert^2 \leq n_k (n_k + 1)~, 
\label{quantumCS}
\end{eqnarray}
see Appendix~\ref{app:CSclass} for more details. To characterize the level of coherence, we shall use the parameter of Ref.~\cite{Campo:2004sz}
\begin{eqnarray}
\delta_k \doteq n_k + 1 - \vert c_k \vert^2 / n_k ~,
\label{delta}
\end{eqnarray}
which belongs to the interval $[0,n_k+1]$. When $\delta_k =0$, one has a maximally entangled squeezed state with zero entropy, and when $\delta_k = n_k + 1 $ one has an incoherent thermal state of maximum entropy. For homogeneous Gaussian states, one also verifies that the entropy is monotonically growing with $\delta_k$. 

The important and nontrivial fact is that $\delta_k = 1$ divides states that are quantum mechanically entangled from states that only possess classical correlations. To show this we recall the notion of separability. A two-mode state is called \textit{separable} when it can be written as a weighted sum of products of two one-mode states, where all weights are positive and can thus be interpreted as probabilities. In this case, the strength of the correlations is more restricted than \eq{quantumCS}. Indeed, one finds $\vert c_k \vert^2 \leq n_k^2$, see Appendix~\ref{app:CSclass}. As a consequence, whenever 
\begin{eqnarray}
\label{nonsep}
n_k^2 < \vert c_k \vert^2 \leq n_k (n_k + 1)~, 
\end{eqnarray}
a homogeneous state is {\it nonseparable}, i.e., so entangled that it cannot be represented as a classically correlated state characterized by probabilities. In terms of $\delta_k$ this criterion is simply given by $\delta_k < 1$.

\subsection{Invariant states and \texorpdfstring{$P$}{P} representation}

Since the states we consider are invariant under the affine group, $n_k$ and $c_k$ are necessarily independent of $k$. We shall nevertheless keep the label $k$ to remind the reader that we work at fixed $k$ and not at fixed $\omega$ as in the next section. Because of the affine group, 
\begin{eqnarray} 
\varphi(P) \doteq \sqrt{k/H} \times \varphi_k(t)~, 
\label{eq:phiPout}
\end{eqnarray}
only depends on $P $, where $\varphi_k(t)$ is the (positive unit norm) \textit{out} mode of \eq{eq:phi_k-decomposition}. The norm of the mode $\varphi$ is fixed by the Wronskian 
\begin{equation}
\label{eq:wronskien}
W(\varphi) = 2 H^2  {\rm Im}(\varphi^* \partial_P  \varphi) = 1~. 
\end{equation}
Using such $\varphi$ and Eqs.~\eqref{Prep2ptfn} and \eqref{eq:phi-Ga}, \eq{eq:acgrennfunction} can be written as
\begin{subequations}
\begin{align}
\left.G_{\rm ac}(P,P)\! \right|_{P \rightarrow 0}=  2\left[2 n_k + 1\right] \vert  \varphi(P) \vert^2 + 4  
{\rm Re} \left (c_k  \varphi^2(P) \right )~, \\
\label{eq:GacP}
=\! \! \iint_0^\infty \! \! \frac{dP_1}{P_1^2} \frac{dP_2}{P_2^2} G_{\rm ret}(P,P_1)G_{\rm ret}(P,P_2) 
N(P_1,P_2)~. 
\end{align}
\end{subequations}
In the second line, the noise kernel of Eq.~\eqref{eq:disskernel}, which is also invariant under the affine group for the set of states we are considering, has been written in the $P$-representation using \eq{Prep2ptfn}. To extract $n_k$ and $c_k$ from the above equations, we need to compute $ G_{\rm ret}$ and $N$. 

Using \eq{Prep2ptfn}, \eq{locEq} reads
\begin{equation}
\label{eq:retardedGf}
 \left[H^2 \partial_P^2\! - \!\! \frac{\gamma(-i P)}{\Lambda  \sqrt{P}} H \partial_P\! \frac{\gamma(i P)}{ \sqrt{P}} \! +\!\! \frac{F^2}{P^2}\right] \!\! G_{\rm ret}(P, P') \!\! = \!\! \delta(P-P') ~.
\end{equation}
The unique (retarded) solution can be expressed as
\begin{equation}
\label{eq:Gret-solution}
\begin{split}
  G_{\rm ret} (P,P') \! =\! 2  \theta(P' - P) {\rm {Im}} \, \left( \tilde \varphi_P \tilde \varphi^\ast_{P'}\right) e^{-\mathcal{I}_P^{P'}} ~,
  \end{split}
\end{equation}
with the optical depth~\cite{Adamek:2008mp},
\begin{equation}
\label{eq:opticaldepth}
 \mathcal{I}_P^{P'} = \int_{P}^{P'}\!dP_1 \frac{\Gamma(P_1) }{ H P_1 }~.
 \end{equation}
Its role is to limit the integrals over $P_1$ and $P_2$ in \eq{eq:GacP} to low values so that $\mathcal{I}_0^P \lesssim 1$. All information about the state for higher values of $P$ is erased by dissipation. In \eq{eq:Gret-solution} we have introduced
\begin{equation}
\label{eq:tildephi}
\tilde \varphi_P \doteq e^{  \mathcal{I}_0^{P'}} \bar \varphi(P)~,
\end{equation}
where $\bar \varphi$ is a homogeneous damped solution of \eq{eq:retardedGf}. By construction, $\tilde \varphi_P$ obeys the reversible (damping free) equation\footnote{
For high values of $P$, the effective dispersion relation is superluminal if $\partial_P (f - \Gamma^2) > 0$, and subluminal if this quantity is negative. The critical case, $f - \Gamma^2 = 0$, gives rise to a relativistic dispersion. In the case where $F^2 - \Gamma^2$ becomes negative, the mode enters an overdamped regime, see Ref.~\cite{Adamek:2008mp}. To avoid the complications this entails, we will only consider $f - \Gamma^2 \geq 0$.}
\begin{equation}
\label{eq:eom-transformed}
 \left[H^2 { P^2}\partial_P^2 + {F^2 - \Gamma^2}\right] \tilde \varphi_P = 0~,
\end{equation}
and is normalized by \eq{eq:wronskien}. Moreover, we impose that it obeys the \textit{out} positive frequency condition, meaning that in the limit $P\to 0$, it asymptotes to the \textit{out} mode $\varphi$ of \eq{eq:phiPout}. Hence, comparing Eq.~(\ref{eq:phi-Ga}) with Eqs.~(\ref{eq:GacP}) and (\ref{eq:Gret-solution}), we find 
\begin{subequations}
\label{eq:n_k,c_k}
\begin{align}
\label{eq:n_k}
n_k \! + \frac12 &= \! \! \iint_0^\infty\!\! \frac{dP_1}{P_1^2}\! \frac{dP_2}{P_2^2} {\rm Re}\left(\tilde \varphi_{P_1} \tilde  \varphi^\ast_{P_2}\right) e^{-\mathcal{I}_0^{P_1}\!-\mathcal{I}_0^{P_2}}\!  N(P_1, P_2)~,\\
\label{eq:c_k}
 c_k &=  \!\iint_0^\infty\!\frac{dP_1}{P_1^2}\! \frac{dP_2}{P_2^2} \tilde \varphi^\ast_{P_1} \tilde \varphi^\ast_{P_2} e^{-\mathcal{I}_0^{P_1}-\mathcal{I}_0^{P_2}}  N(P_1, P_2)~.
 \end{align}
\end{subequations}
These central equations establish how the environment noise kernel $N$ fixes the late time mean occupation number and the strength of the correlations. 

We now compute $N$. When $u$ is freely falling, the rescaled field $\hat \Psi^0_q$ of Eq.~\eqref{eq:psirenorm} is a dense set of independent harmonic oscillators of constant frequency $\Omega_q= \pi \Lambda |q|$, one at each $z$. The frequency is constant because we set $c_\psi = 0$ in the action for $\psi$, see the discussion after \eq{eq:covaction}. It implies that the positive frequency mode functions are the standard $e^{-i \Omega_q t} / \sqrt{2 \Omega_q}$, and that the state of these oscillators remains unaffected by the expansion of the universe. Hence $T_\psi$, the temperature of the environment, is not redshifted. 

We here wish to recall that for relativistic (and dispersive) fields, the vacuum state of zero temperature is the only stationary state which is Hadamard~\cite{Busch:2012ne}. Hence, for these fields, the temperature is fixed to zero. This is not the case in our model where any temperature $T_\psi$ is acceptable. In what follows, we shall thus treat $T_\psi$ as a free parameter, and work with homogeneous thermal states. This means that the expectation value of the anticommutator of $\hat \psi^0_q$ is given by
\begin{equation}
\begin{split}
\label{eq:NofDelta}
{\rm Tr} \left (\hat \rho_{\rm tot} \{\hat \psi_q^{0}(\mathsf x),\hat \psi_{q'}^{0}(\mathsf x')\}\right ) &=\frac{ \delta(z - z')}{\sqrt{a(t) a(t')}} \delta(q-q') \\
&\quad \times \coth\frac{\Omega_q}{2 T_\psi}\ \frac{ \cos\left ( \Omega_q \Delta t \right )}{\Omega_q}~.
\end{split}
\end{equation}
The factor $ \coth({\Omega_q}/{2 T_\psi}) =2 n^\Psi_{ q} + 1 $ is the standard bosonic thermal distribution. The prefactor $\delta(z - z')/\sqrt{a(t) a(t')}$ comes from the facts that $\hat \Psi^0_q$ of Eq.~\eqref{eq:psirenorm} is a dense set of independent oscillators, and that $a(\tau,z)$ reduces here to the scale factor ${a(t)}$. To get $N$ of \eq{eq:disskernel} one should differentiate the above and integrate over $q$. The integration gives a distribution which should be understood as Cauchy principal value,
\begin{equation}
\begin{split}
\label{eq:NofDelta2}
\iint dq dq' \nabla_{u} \nabla_{u'} {\rm Tr} &\!\left (\hat \rho_{\rm tot} \{\hat \psi_q^{0}(\mathsf x),  \hat \psi_{q'}^{0}(\mathsf x')\}\right )= - \frac{ \delta(z - z')}{\sqrt{a(t) a(t')}} \\
&\quad \quad \times \frac{2 T_\psi }{H \Lambda }  \frac{\partial}{\partial{ \Delta t}}\, \mathtt{P.V.}\coth \!\left(\pi T_\psi \Delta t\right)~.
\end{split}
\end{equation}
To be able to re-express \eq{eq:NofDelta2} in the $P$-representation, it is necessary to verify that it is invariant under the affine group. This is easily done using notations of the Appendix~\ref{App:Prep}. One verifies that the first factor simply equals $\delta(\Delta_2)$, whereas the second line is only a function of $\Delta_1$. Taking into account the derivatives of Eq.~\eqref{eq:disskernel}, in the $P$-representation, the noise kernel at temperature $T_\psi$ reads
\begin{equation}
 \label{eq:noisekernel}
 \begin{split}
N (P, P') = &- \frac{\gamma(i P) \gamma(-i P')}{\Lambda } {2 T_\psi}{\sqrt{P P'}} \\
&\quad \quad \frac{\partial}{\partial{\ln\frac{P'}{P}}}\, \mathtt{P.V.}\coth \!\left(\frac{\pi T_\psi}{H} \ln\frac{P'}{P}\right)~.
 \end{split}
\end{equation}
The symbol $\mathtt{P.V.}$ indicates that when evaluated in the integrals of Eq.~\eqref{eq:n_k,c_k}, the nonsingular part should be extracted using a Cauchy principal value prescription on $\ln(P'/P) = H (t- t')$. 

In the high-temperature limit, the double integrals of \eq{eq:n_k,c_k} can be evaluated analytically because $N$ effectively acts as a Dirac delta function. Instead, when working with an environment in its ground state, or at low temperature $T_\psi$, we are not aware of analytical techniques to evaluate these integrals. Hence, to study the impact of dissipation on coherence in (near) vacuum states, we shall numerically integrate Eqs.~\eqref{eq:n_k,c_k}.

\subsection{Numerical Results}

In the forthcoming numerical computations, for simplicity, we work with
\begin{equation}
\label{eq:numdispersion}
f = \frac{P^4}{\Lambda^2}, \quad \Gamma = g^2 \frac{P^2}{2 \Lambda}~,
\end{equation}
which contain the same ultraviolet momentum scale $\Lambda$. The dimensionless coupling $g^2$ controls the relative importance of dispersive and dissipative effects. In the limit $g^2 \rightarrow 0$, we get the quartic superluminal dispersion studied in Refs.~\cite{Macher:2008yq,Busch:2012ne}. The critical coupling $g^2_{\rm crit}= 2$, greatly simplifies the calculations, since $f - \Gamma^2 = 0$ guarantees that $\tilde \varphi_P$ obeys a relativistic equation, see \eq{eq:eom-transformed}. 

Using a numerically stable procedure for the Cauchy principal values like in Ref.~\cite{Adamek:2008mp}, we compute $n_k$ and $c_k$ of Eq.~\eqref{eq:n_k,c_k} in the parameter space $\Lambda$, $g^2$, $m^2$, and $T_\psi$. Since all physical effects only depend on dimensionless ratios, we present the numerical results in terms of $\mu = m/H$, $\lambda = \Lambda/H$, and $\vartheta = T_\psi/H$. 

\begin{figure}[tb]
\vspace{-15pt}
\centerline{\includegraphics[width=0.95\columnwidth]{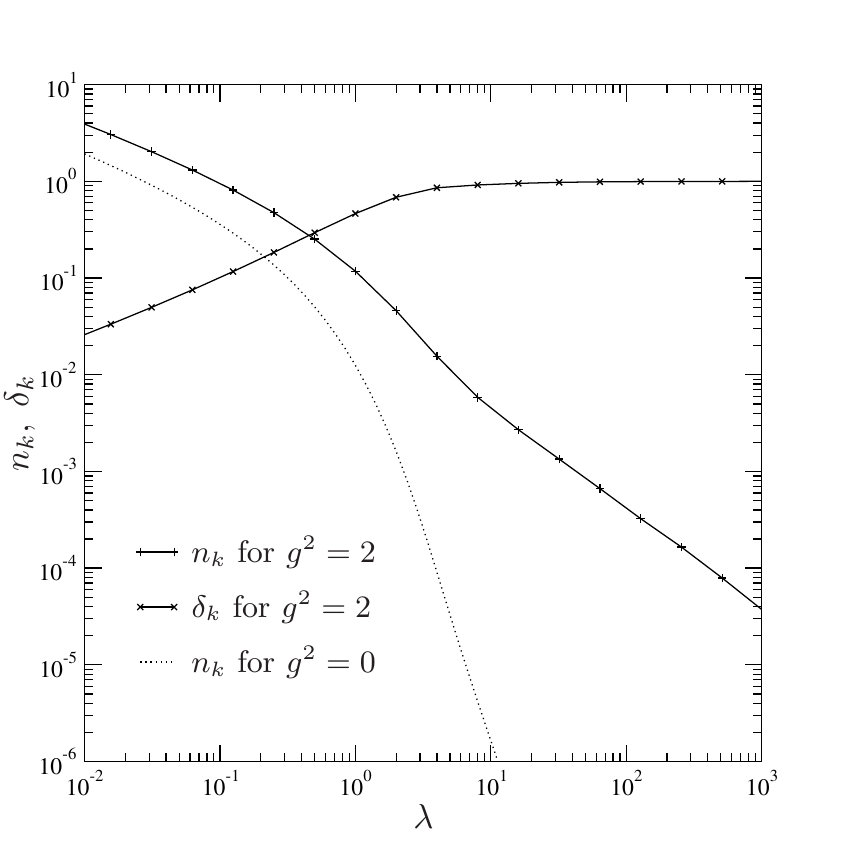}}
\caption{\label{fig:massless-critical} \small
Numerical values for $n_k$ and $\delta_k$ for a massless field with \textit{critical} damping $g= g_\mathrm{crit}$ and quartic superluminal dispersion at the energy scale $\Lambda = H \lambda$. For comparison, we have represented by a dotted line the $n_k$ of the quartic dispersive field (in which case $\delta_k = 0$ identically). Surprisingly, the state is nonseparable, $\delta_k < 1$, for all values of $\lambda$. Moreover, $\delta_k$ decreases when dissipation increases. }
\end{figure}

\subsubsection{Massless critical case}
 
 \begin{figure*}[tb]
\begin{minipage}{0.45\linewidth}
\centerline{\includegraphics[width=0.95\columnwidth]{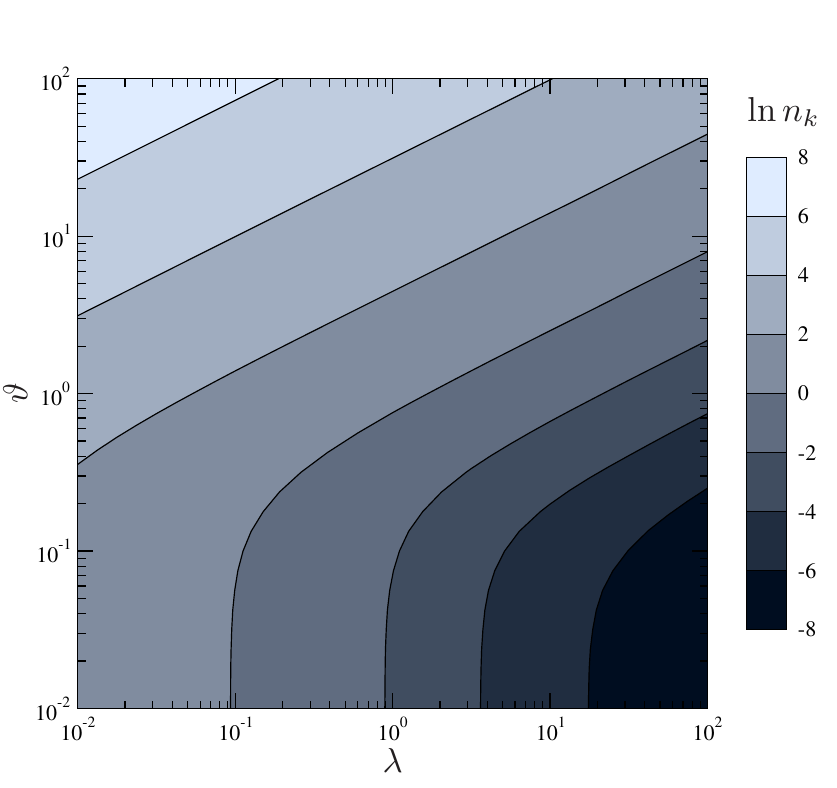}}
\end{minipage}
\hspace{0.05\linewidth}
\begin{minipage}{0.45\linewidth}
\centerline{\includegraphics[width=0.95\columnwidth]{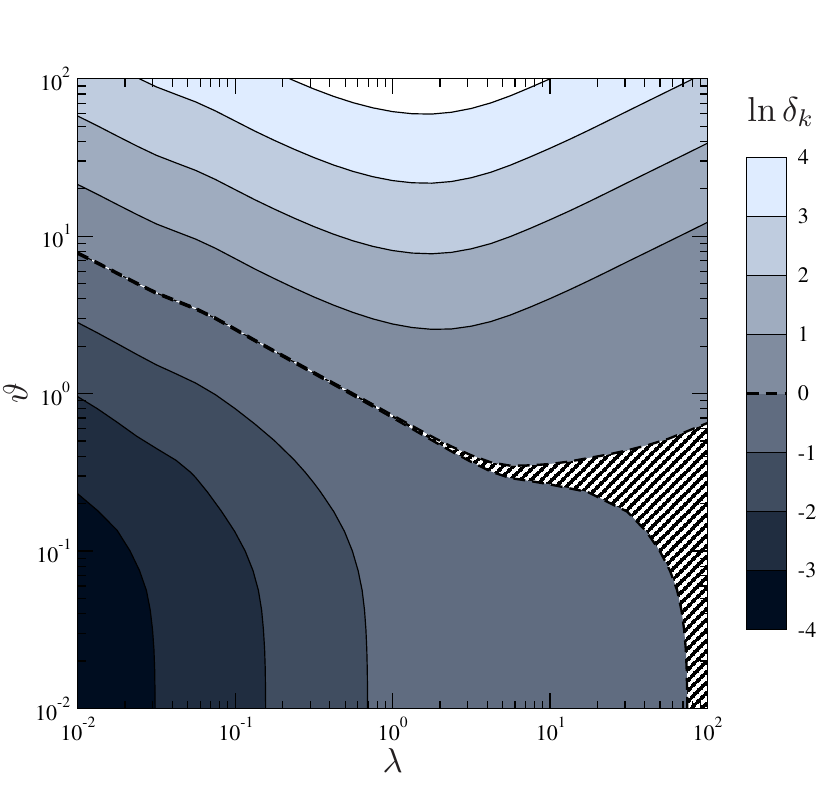}}
\end{minipage}
\caption{\label{fig:massless-T} \small
Contour plots of $\ln n_k$ and $\ln \delta_k$ for a massless field with critical coupling $g= g_\mathrm{crit}$ in the parameter space $(\lambda = \Lambda/H,~\vartheta = T_\psi/H)$ . At low temperatures, for $\vartheta = T_\psi/H \lesssim 1/10$, $n_k$ and $\delta_k$ barely depend on $\vartheta$. On the contrary, for high temperatures, $\vartheta \gtrsim  1$, $n_k$ scales as $n_k \propto \vartheta \lambda^{-1/2}$ whereas $\delta_k$ scales as $\delta_k \propto \vartheta \lambda^{-1/2}$ for $\lambda \gtrsim 1$, and $\delta_k \propto \vartheta \lambda^{1/2}$ for $\lambda \lesssim 1$. The hatched region indicates the numerical uncertainty about the threshold value $\delta_k = 1$ found when $n_k \ll 1$. }
\end{figure*}

We begin with the massless case ($\mu^2 = 0$) and with $g=g_{\rm crit}$. Then Eq.~(\ref{eq:eom-transformed}) is particularly simple since the rescaled mode $\tilde \varphi$ of \eq{eq:tildephi} reduces \textit{for all} $P$ to the \textit{out}-mode $ \varphi_P = {e^{i P/H}}/{\sqrt{2 H}}$. In this we recover the conformal invariance of the massless field in two dimensions. There usually would be no particle production when it propagates in de Sitter space, however, the conformal invariance being broken by dissipation, pair-creation will take place. 

In Fig.~\ref{fig:massless-critical} we present $n_k$ and $\delta_k$ when the environment is in its ground state ($T_\psi = 0$). For comparison, we also show $n_k$ for quartic dispersion ($g^2 = 0$) which can be computed analytically in the Bunch-Davies vacuum~\cite{Macher:2008yq}. For $\lambda \to \infty$ the number of particles goes to zero as $1/ \lambda$, as is expected since conformal invariance is restored in this limit. Despite dissipation, we find that $\delta_k < 1$ for all values of $\lambda$. This indicates that the state is always nonseparable in the two-mode $k$ basis. In addition, contrary to what might have been expected, the two-mode entanglement is stronger for smaller values of $\lambda$, i.e., stronger dissipative effects. The reason for this has to be found in the fact that $\lambda$ also sets the scale where conformal invariance is broken. 

Let us now turn to the effects of the environment temperature $T_\psi$. Figure~\ref{fig:massless-T} shows contour plots of $n_k$ and $\delta_k$ for a massless field with \eq{eq:numdispersion}, again for $g= g_{\rm crit}$. In the limit $\lambda \to \infty$, we observe that $n_k\to 0$ irrespectively of the value of $T_\psi$. This establishes that there is a robustness of the relativistic result in the limit $\lambda \to \infty$ which generalizes that found for dispersive fields, see e.g., Ref.~\cite{Macher:2008yq}. Moreover, in the high-temperature limit, Eqs.\eqref{eq:n_k,c_k} can be evaluated analytically to give
\begin{subequations}
\label{eq:massless-highT}
\begin{align}
\label{eq:n_massless-highT}
n_k + \frac{1}{2} &\sim \frac{\sqrt{\pi} \vartheta}{\sqrt{\lambda}}~,\\
\label{eq:d_massless-highT}
\delta_k &\sim \frac{\sqrt{\pi} \vartheta}{\sqrt{\lambda}} \left(1 - \frac{1 + \mathrm{erfi}^2\sqrt{\lambda}}{e^{2\lambda}}\right)~,
 \end{align}
\end{subequations}
where $\mathrm{erfi}$ is the imaginary error function. We compared the corresponding contours with the numerical ones shown in Fig.~\ref{fig:massless-T} and found that they are practically indistinguishable for $\vartheta > 10$.

When considering the effects of $T_\psi$, we observe two regimes. At low temperature ($\vartheta \ll 1$), $n_k$ and $\delta_k$ only depend on $\lambda$ and are basically given by the zero temperature limit shown in Fig.~\ref{fig:massless-critical}. However, at large temperature ($\vartheta \gg 1$), they depend on $\lambda$ and $\vartheta$ according to Eqs.~\eqref{eq:massless-highT}. As expected, the strongest signatures of quantum entanglement, $\delta_k \ll 1$, are found in the region where the breaking of conformal invariance is large (and hence pair-creation is active) and when the environment temperature is small, so that the spontaneous pair-creation events are not negligible with respect to thermally induced events. On the other hand, when the temperature is large, the final state is separable since $\delta_k \gg 1$. In Fig.~\ref{fig:massless-T} (right panel) we see that the threshold case $\delta_k = 1$ is approximatively given by $\vartheta \sim \lambda^{-1/2}$ for $\lambda \lesssim 1$. The hatched region for $\lambda \gtrsim 10$ represents the numerical uncertainty in the region where $n_k$ is much smaller than 1.

\begin{figure*}[tb]
\begin{minipage}{0.45\linewidth}
\centerline{\includegraphics[width=0.95\columnwidth]{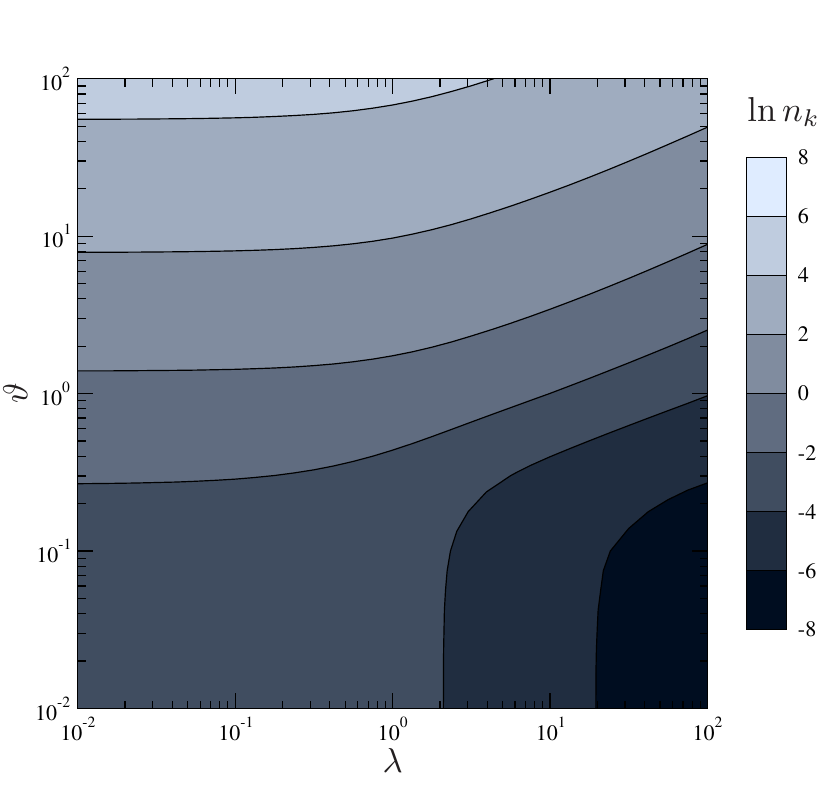}}
\end{minipage}
\hspace{0.05\linewidth}
\begin{minipage}{0.45\linewidth}
\centerline{\includegraphics[width=0.95\columnwidth]{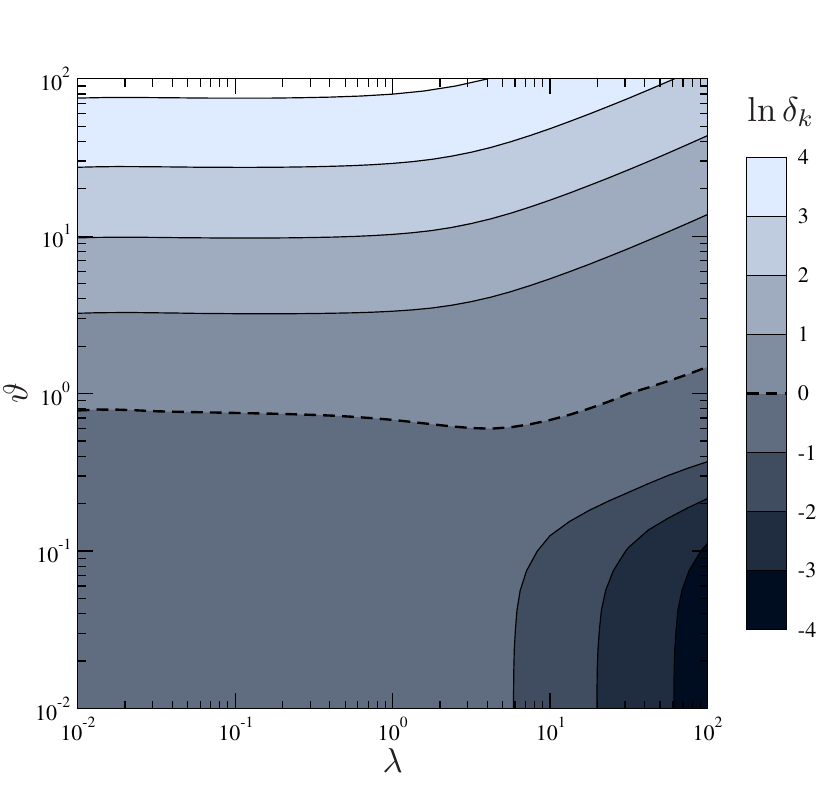}}
\end{minipage}
\caption{\label{fig:massive-T} \small
Contour plots of $\ln n_k$ and $\ln \delta_k$ for a massive field ($\mu^2 = 5/4$) with critical coupling $g^2_\mathrm{crit} = 2$ in the parameter space $(\lambda = \Lambda/H,~\vartheta = T_\psi/H)$. As in Fig.~\ref{fig:massless-T}, for low temperature $\vartheta \lesssim 0.1$, $n_k$ and $\delta_k$ are independent of the temperature. Instead for $\vartheta \gtrsim 1$ and $\lambda \gg 1$, $n_k$ and $\delta_k$ scale both as $\vartheta \lambda^{-1/2}$. }
\end{figure*}

\subsubsection{Massive fields}

We note that the massless case $\mu^2 = 0$ is an isolated point in the mass spectrum: a well-defined notion of \textit{out}-quanta requires either $\mu^2 = 0$ or $\mu^2 > 1/4$. In the latter case, the asymptotic \textit{out}-modes with positive frequency (see, e.g., Appendix B of Ref.~\cite{Macher:2008yq}) are given by
\begin{equation}
 \varphi_P = \sqrt{\frac{\pi}{2 \sinh \pi \tilde{\mu}}} \frac{\sqrt{P}}{H} J_{i \tilde{\mu}} (P / H)~,
\end{equation}
where $\tilde{\mu} \doteq \sqrt{\mu^2 - {}^1\!/_4}$ and $J$ denotes the Bessel function of the first kind.

Figure~\ref{fig:massive-T} shows the contour plots of $n_k$ and $\delta_k$ for a massive field with $\mu^2 = 5/4$ and $g= g_\mathrm{crit}$, in the same parameter space ($\lambda$, $\vartheta$) as in Fig.~\ref{fig:massless-T}. The case of a Lorentz-invariant field in the Bunch-Davies state is recovered in the limit $\lambda \rightarrow \infty$, $\vartheta \rightarrow 0$. Now conformal invariance is already broken by the mass term and therefore $n_k$ remains nonzero in this limit.

At zero temperature, the strongest entanglement (lowest $\delta_k$) is found at large values of $\lambda$, i.e., weak dissipation. This was expected, since dissipation reduces the strength of correlations. However, as in the massless case, the threshold of separability $\delta_k = 1$ is not crossed. 

When increasing the environment temperature $T_\psi$, we see that the strength of correlation is reduced, and separable states are found. The nonseparability criterion $\delta_k < 1$ is therefore only met either when $T_\psi$ is smaller than the Gibbons-Hawking temperature $T_{\rm GH} = H/2\pi$, or when the coupling to the environment is sufficiently weak. Notice also that the behavior at high temperature can again be obtained analytically, the integrals over the Bessel functions becoming hypergeometric functions.

\subsubsection{Role of \texorpdfstring{$g$}{g} in the underdamped regime}

\begin{figure*}[tb]
\begin{minipage}{0.45\linewidth}
\centerline{\includegraphics[width=0.95\columnwidth]{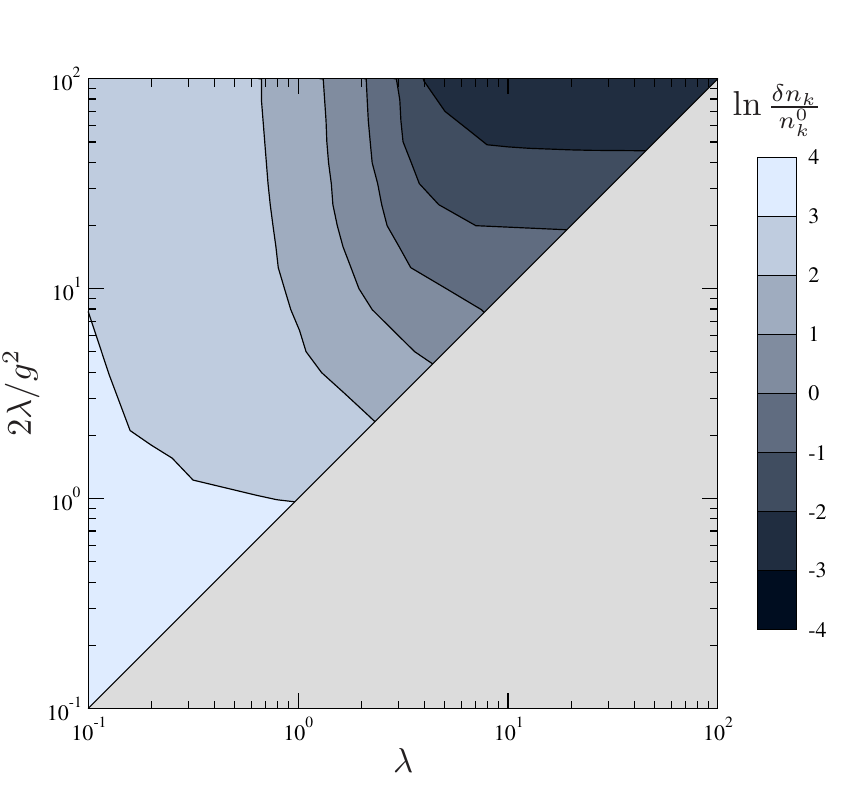}}
\end{minipage}
\hspace{0.05\linewidth}
\begin{minipage}{0.45\linewidth}
\centerline{\includegraphics[width=0.95\columnwidth]{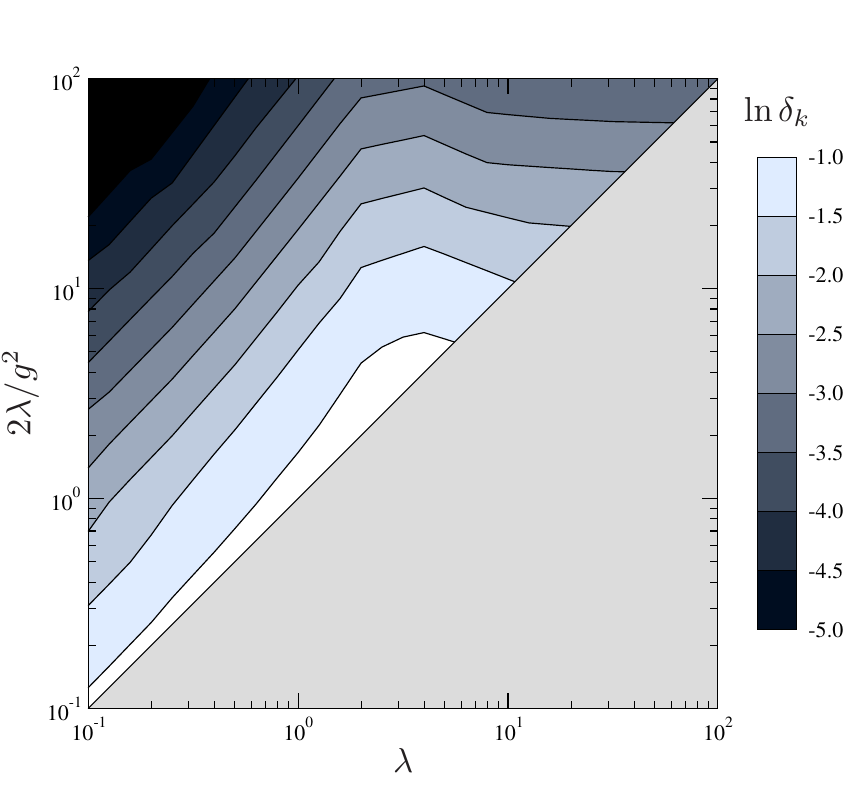}}
\end{minipage}
\caption{\label{fig:massive-vac} \small
Contour plots of $\delta n_k / n_k^0$ and $\delta_k$ for a massive field ($\mu^2 = 5/4$) in the \textit{underdamped} regime $g^2 \leq g_\mathrm{crit}^2$. The environment is in its ground state ($\vartheta = 0$) and the two axes are the dispersive scale $\lambda$ and the dissipative one $2 \lambda / g^2 \geq \lambda$.}
\end{figure*}

It is also interesting to consider the role of the coupling $g$, see \eq{eq:numdispersion}. As $g^2$ approaches zero, the dissipative scale $2\Lambda/g^2$ is moved deeper into the UV with respect to the dispersive scale which is fixed by $\Lambda$. In the limit $g^2 \rightarrow 0$, the field becomes purely dispersive and $n_k$, $\delta_k$ can be computed analytically~\cite{Macher:2008yq} in the Bunch-Davies vacuum. For $g^2 < 2$ the mode is underdamped. In this case, the solutions to Eq.~(\ref{eq:eom-transformed}) which correspond to asymptotic \textit{out}-modes of positive frequency are given by, see Appendix B of Ref.~\cite{Macher:2008yq}, 
\begin{equation}
\tilde \varphi_P = \frac{e^{-\pi \tilde{\mu} / 4}}{\sqrt{P}} \sqrt{\frac{\tilde{\lambda}}{\tilde{\mu}}} M_{i \frac{\tilde{\lambda}}{2}, i \frac{\tilde{\mu}}{2}} \!\left(-i\frac{P^2}{2 \tilde{\lambda} H^2}\right)~,
\end{equation}
where $\tilde{\lambda} \doteq \lambda / \sqrt{4 - g^4}$ and $M$ is a Whittaker function defined in Ref.~\cite{Abramowitz}.

Figure~\ref{fig:massive-vac} shows contour plots of $\delta n_k / n_k^0 \doteq (n_k - n_k^0) / n_k^0$ (where $n_k^0$ is the number of particles without dispersion and dissipation) and $\delta_k$ for a massive field in the underdamped regime. Here, we set $T_\psi = 0$, and plot the results in the parameter space spanned by the two (dimensionless) ultraviolet scales: $\lambda$ which characterizes dispersion, and $2 \lambda / g^2$ which is the UV scale of dissipation. The latter is larger than the former in the underdamped regime. The grey areas therefore correspond to the overdamped regime which we did not study. 

In the weak dispersive/dissipative regime $\lambda \gtrsim 10$, it is evident that $\delta n_k$ and $\delta_k$ are both dominated by dissipative effects. For the latter, this is because dispersion alone does not lead to decoherence. For the deviation $\delta n_k$, this follows from the fact that dispersion gives an exponentially small correction to the pair creation process (see Ref.~\cite{Macher:2008yq}), while the corrections due to dissipation are only algebraically small. As a result, the hierarchy of scales does not directly fix the importance of the respective effects.

On the other hand, when dispersion is strong ($\lambda \lesssim 1$) the pair creation process is basically governed by dispersive effects. The correction to the particle number due to dissipation is very small (compared to the dispersive correction). One can also observe that the degree of two-mode entanglement is then basically governed by the separation between the two scales $g^2$, i.e., $\delta_k$ is determined by the strength of dissipation \textit{at the dispersive threshold}, $\left (\Gamma/P\right )\vert_{ P = \Lambda}$.

\section{Stationary picture}
\label{sec:statio}

In the absence of dispersion/dissipation, it is well known that the Bunch-Davies vacuum is a thermal (KMS) state at the Gibbons-Hawking temperature $T_{\rm GH} = H/2\pi$~\cite{birrell1984quantum}. It is also known that this is the temperature seen by any inertial particle detector, and that this is closely related to the Unruh effect found in Minkowski space, and to the Hawking radiation emitted by black holes~\cite{Brout:1995rd}. In the presence of dissipation, while the stationarity of the state of $\phi$ is {\it exactly} preserved when the state of the environment is invariant under the affine group, the thermality of the state is {\it not} exactly preserved. This loss of thermality, which generalizes what was found for dispersive fields~\cite{Busch:2012ne}, questions the status of black hole thermodynamics when Lorentz invariance is violated~\cite{Dubovsky:2006vk,Eling:2007qd,Jacobson:2008yc}.

\subsection{Loss of thermality}

\begin{figure*}[tt]
\begin{minipage}[t]{0.45\linewidth}
\centerline{\includegraphics[width=1\columnwidth]{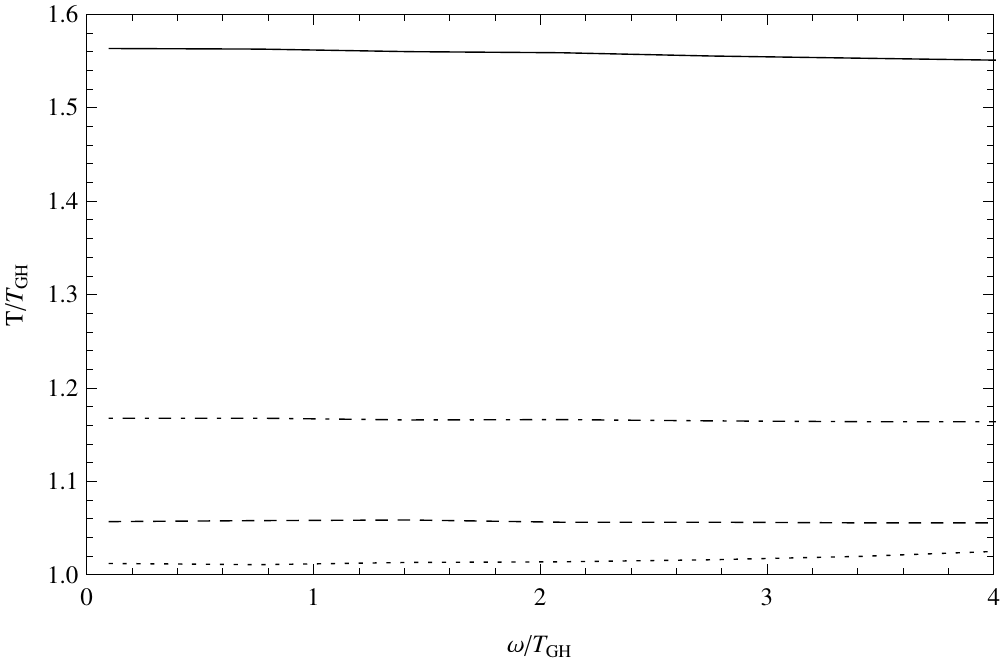}}
\caption{\label{fig:Tofom} \small
Plot of the ratio $T_\om/T_{\rm GH}$ as a function of $\omega/ T_{\rm GH}$ for various values of $\lambda$. We work with a massless field with $g= g_{\rm crit}$, for a detector localized in the center of the patch ($X=0$), and with $T_\psi= 0$. The values of $\lambda$ are $1$ (continuous), $3$ (dot-dashed), $5$ (dashed), and $10$ (dotted.)}
\end{minipage}
\hspace{0.05\linewidth}
\begin{minipage}[t]{0.45\linewidth}
\centerline{\includegraphics[width=1\columnwidth]{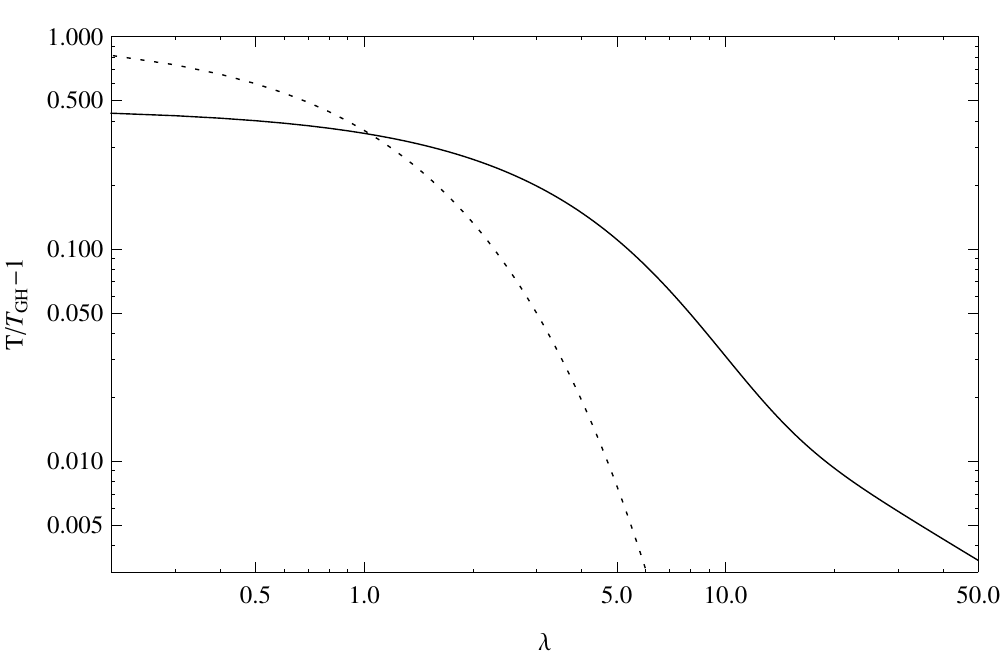}}
\caption{\label{fig:logToflambda} \small
Plot of $T_0/T_{\rm GH}-1$ in logarithmic scales as a function of $\lambda$, where $T_0$ is the low-frequency temperature of massless fields with $g= g_{\rm crit}$ and when $\vartheta = 0$ ($\psi$-vacuum). We have also represented by a dotted curve the same quantity evaluated without dissipation when the state is the Bunch-Davies vacuum.}
\end{minipage}
\end{figure*}

To probe the stationary properties of the state, we consider the transition rates of particle detectors at rest with respect to the orbits of $K_t$. This means that the detector is located at fixed $H|X| < 1$ in the coordinates of \eq{eq:PG}. In this case, the two-point functions only depend on $t-t'$ and can be analyzed at fixed $\omega = i \partial_t\vert_X $, see \eq{eq:fouriert}. (The above restriction on $X$ simply expresses that the trajectory be timelike.)

The transition rates are, up to an overall constant, given by Fourier transforms of the Wightman function $G_W$~\cite{Brout:1995rd}. The rates then determine $n_\omega(X)$, the mean number of particles of frequency $\omega > 0$ seen by a detector located at $X$, through
\begin{equation}
\label{eq:nofG}
\begin{split}
\frac{n_\omega (X)}{n_\omega (X)+1} = \frac{G_{W}^\omega(X,X) }{ G_{W}^{-\omega}(X,X) }~. 
\end{split}
\end{equation}
To study the deviations with respect to the Gibbons-Hawking temperature $T_{\rm GH} = H/2 \pi$, we introduce the temperature function $T_\omega (X)$ defined by
\begin{equation}
\label{eq:temperature}
\frac{n_\omega (X)}{n_\omega (X)+1} = \ep{- \omega/ T_\omega (X)}~.
\end{equation}
It gives the effective temperature seen by the detector, and reduces to the standard notion when it is independent of $\omega$. In the absence of dispersion and dissipation, $T_\omega (X) = T_{\rm GH}$ for all values of $\omega$, which means that the Tolman law is satisfied~\cite{Busch:2012ne}. 

In the following numerical computations, for simplicity, we work at $X=0$ with an inertial detector, with $g= g_{\rm crit}$, $m=0$, and $T_\Psi =0$. Since the calculation of the commutator of $\phi$ is much faster and more reliable than that of the anticommutator, instead of using Eq.~\eqref{eq:nofG}, $n_\omega$ shall be computed with
\begin{equation}
\begin{split}
 n_\omega (X)= \frac{G_{W}^\omega(X,X) }{ G_{\rm c}^{\omega}(X,X) }~. 
\end{split}
\end{equation}
The denominator is expressed using Eq.~\eqref{eq:GcfromGret}. The numerator is obtained from Eqs.~(\ref{eq:GacXXP}) and~(\ref{eq:noisekernel}) with $T_\psi \to 0$. In addition, the principal value is replaced by a prescription for the contour of $\ln P/P' = H t $ to be in the upper complex plane. In this we recover the fact that when the anticommutator in the vacuum is $\mathtt{P.V.} (1/t)$, the corresponding vacuum Wightman function is $1/(t-i \epsilon)$.

In Fig.~\ref{fig:Tofom}, we plot the ratio $T_\omega/T_{\rm GH}$ as a function of $\omega$ for various values of $\lambda$, and for $T_\psi = 0$. We first observe that $T_\omega$ is constant for all frequencies from zero to a few multiples of $T_{\rm GH}$. Hence, the Planckian character of the state is, to a high accuracy, preserved by dissipation, as was found in the presence of dispersion~\cite{Macher:2009tw,Finazzi:2010yq,Busch:2012ne}. For higher frequencies, i.e., $\omega/T_{\rm GH} >  4$, we were not able to study $T_\omega$ with sufficient accuracy because of the numerical noise associated to $n_\omega  < 0.01$. As in the dispersive case, we expect that the temperature function $T_\omega$ is modified for $\omega \gtrsim \Lambda$. 

Secondly, when $\lambda$ is smaller than 5, i.e., when dissipation is strong, we observe that the temperature is significantly (more than $5\%$) larger than $T_{\rm GH}$. These deviations are further studied in Fig.~\ref{fig:logToflambda}, where we plot the deviations of $T_0$, the low-frequency effective temperature, with respect to $T_{\rm GH}$ as a function of $\lambda$. We observe that the deviation due to dissipation asymptotically follows 
\begin{equation}
\frac{T_0}{T_{\rm GH}}-1 \underset{\lambda \to \infty}\sim (6 \lambda)^{-1} ~. 
\end{equation}
This law has been verified up to $\lambda = 10^3$. It has to be compared with the deviation due to quartic dispersion studied in Ref.~\cite{Busch:2012ne}. This deviation is represented by the dotted curve, and scales as ${T_0^{\rm disp}}/{T_{\rm GH}}-1 \sim \ep{-\pi \lambda/4}$. In other words, the deviation due to (quadratic) dissipation decreases much slower than that due to (quartic) superluminal dispersion. The important lesson for black hole thermodynamical laws is that ultraviolet dispersion and dissipation both destroy the thermality of the state. This lends support to the claim that Lorentz invariance is somehow necessary for these laws to be satisfied.

\subsection{Asymptotic correlations among right movers}
\label{Asc}

As explained in Sec.~\ref{sec:homoentangle}, at late time, the $\phi$ field decouples from its environment. This allows to use the relativistic \textit{out} basis at fixed $k$ to read out the state of $\phi$. Alternatively, one can also use an \textit{out} basis formed with stationary modes with fixed frequency $\omega$. Indeed, at fixed $\omega$, the momentum $P_\om \sim |\omega/X| \to 0$ at large $|X|$, and dispersive effects are negligible. Hence $\hat \phi_\omega(X)$, the stationary component of the field operator, decouples from the environment at large $|X|$, and can be analyzed using relativistic modes. As we shall see, this new \textit{out} basis is {\it not} trivially related to the homogeneous one used in Sec.~\ref{sec:homogen} because it encodes thermal effects at the Gibbons-Hawking temperature. Hence the covariance matrix of the new \textit{out} operators will depend on $n_k$ and $c_k$ of Eq.~\eqref{eq:n_k,c_k}, but also on these thermal effects. At this point we need to explain why we are interested in expressing in a different basis a state which is fully characterized by $n_k$ and $c_k$. The main reason comes from black hole physics. As shall be discussed in the next section, when certain conditions are met, the results of this section apply to the Hawking radiation emitted by dissipative fields. 

To compute the covariance matrix in the new basis, we recall some properties of the relativistic massless field in de Sitter. First, because of conformal invariance, the field operator splits into two sectors which do not mix, one for the right-moving $U$ modes with $\bk > 0$, and the other for the left-moving $V$ modes with $\bk < 0$. In addition, in de~Sitter, the time-dependence of all homogeneous modes can be expressed through $\varphi(P)$ of Eq.~\eqref{eq:phiPout}, which here reduces to
\begin{equation}
\begin{split}
\varphi(P) =   \ep{i P/H} /\sqrt{2 H}~,
 \end{split}
\end{equation}
where $P > 0$. This mode has a unit positive Klein-Gordon norm, as can be verified using the Wronskian condition of Eq.~\eqref{eq:wronskien}.
 
We introduce an intermediate basis constructed with the stationary \enquote{Unruh} modes $\varphi_\om$~\cite{Unruh:1976db}. In the $P$ representation, they can be written as~\cite{Parentani:2010bn}
\begin{equation}
\begin{split}
\label{eq:unruhmodedef}
 \varphi^{\omega} = (P/H)^{-i \omega/H -1} \times \varphi(P) ~. 
 \end{split}
 \end{equation}
They form an orthonormal and complete mode basis if $\omega \in ]- \infty, \infty[$. The spatial behavior of the $U$ modes is given by
\begin{equation}
\begin{split}
\label{varphiX}
\varphi_U^{\omega }(X) = \int_0^\infty \frac{d\bP}{H \sqrt{2\pi}} e^{i \bP X}  \varphi^{\omega}(P)~.
\end{split}
\end{equation}
We now introduce the alternative \textit{out} basis formed of stationary modes which are localized on either side of the horizons, henceforth called $R$ and $L$ modes. They behave as Rindler modes in Minkowski space. For $U$-modes, the horizon is located at $HX = - 1$, and these modes are
\begin{equation}
\label{RLm}
\begin{split}
\chi_U^{\omega, R } (X) &= \theta(1+HX)  \frac{(1+HX) ^{i\omega/H}}{\sqrt{2\omega}}~,\\
(\chi_U^{-\omega, L } (X))^* &= \theta(-1 - HX) \frac{(-1- HX) ^{i\omega/H}}{\sqrt{2\omega}}~,\\
\end{split}
\end{equation} 
where $\om > 0$. The first has a positive norm, while the second has a negative one. They are easily related to the Unruh mode by computing \eq{varphiX}. Indeed, for $\om > 0$, one gets
\begin{equation}
\begin{split}
\label{eq:UnruhRLchange}
\varphi_U^{\omega }  =  \alpha_\omega^H  \chi_U^{\omega, R } +   \beta_\omega^H (\chi_U^{-\omega, L } )^* ~,
\end{split}
\end{equation} 
where coefficients $\alpha_\omega^H$ and $\beta_\omega^H$ are the standard Bogoliubov coefficients leading to the Gibbons-Hawking temperature $H/2\pi$. They obey $\abs{\beta_\omega^H/\alpha_\omega^H} = e^{ - \pi \om/H}$. Asymptotically in the future and in space, the $U$ part of the field operator can thus be expressed as
\begin{subequations}
\label{eq:operatorsphi}
\begin{align}
\label{eq:operatorphihomo}
\hat \phi_U (\mathsf x) = \int_0^\infty d \bk & \{ \hat a_{\bk} \, e^{i \bk z} \, \varphi_k(t) + h.c. \}\\ 
\label{eq:operatorphiunruh}
 = \int_{-\infty}^{\infty} d\om  &\{ \hat a_{U}^{ \om} \, e^{- i \om t} \, \varphi_U^{\omega }(X)  + h.c. \}\\
\label{eq:operatorphiout}
 = \int_{0}^{\infty} d\om  &\{ \hat a_{U, R}^{ \om} \, e^{- i \om t} \, \chi_{U, R}^{\omega }(X)  \\ 
 \nonumber
& + \hat a_{U, L}^{ -\om \, \dagger} \, e^{- i \om t} \, (\chi_{U, L}^{-\omega }(X))^* + h.c. \} 
\end{align}
\end{subequations}
The $V$ part possesses a similar decomposition, and the $V$ modes are obtained from the $U$ ones by replacing $X \to-X$, and $R \to L$. The $\chi_V$ modes are thus defined on either side of $HX = 1$. 

Using the above equations, the Unruh and the Rindler-like operators of frequency $\vert \om \vert$ are related by
\begin{equation}
\label{eq:changeofbase}
\begin{split}
\left (
\begin{array}{l}
\hat a_{U, R}^\om\\
\hat a_{U, L}^{ -\om \, \dagger}\\
\hat a_{V, L}^\om\\
\hat a_{V, R}^{ -\om \, \dagger}
\end{array} \right )=  \left ( \begin{array}{cccc}
\alpha_\omega^H&\beta_\omega^{H\, *}&0&0\\
\beta_\omega^{H}& \alpha_\omega^{H\, *}&0&0\\
0&0&\alpha_\omega^H&\beta_\omega^{H\, *}\\
0&0&\beta_\omega^{H}& \alpha_\omega^{H\, *}\\
\end{array}\right ) \times \left (
\begin{array}{l}
\hat a_U^\om\\
\hat a_U^{- \om\,\dagger}\\
\hat a_V^\om\\
\hat a_V^{- \om\,\dagger}\\
\end{array} \right )~.
\end{split}
\end{equation}
We considered both $U$ and $V$ modes because our aim is to compute the covariance matrix of the $R$ and $L$ operators in terms of $n_k$ and $c_k$ of Eq.~\eqref{eq:n_k,c_k}, where $c_k$ mixes $U$ and $V$ modes. To do so, we first compute the covariance matrix of the Unruh operators. When working with states that are invariant under the affine group, $n_k$ and $c_k$ of Eq.~\eqref{eq:n_k,c_k} are independent of $k$. This implies that the covariance matrix of the Unruh operators is independent of $\om$. Indeed, using 
\begin{equation}
\label{eq:akaomega}
\hat a_U^\omega = \int_0^\infty \frac{d \bk}{H}   \left (\frac{k}{H}\right )^{ i\omega /H -1/2} \hat a_\bk~, 
\end{equation} 
which follows from the Fourier transforms Eqs.~\eqref{eq:operatorphihomo} and~\eqref{eq:operatorphiunruh}, one verifies that the independence of $k$ implies that of $\omega$. As a result, introducing $V_\omega^\dagger = \left ( \hat a_U^{\dagger\, \omega}, \hat  a_U^{-\omega} ,\hat a_V^{\dagger\, \omega},  \hat a_V^{-\omega} \right )$, the covariance matrix of Unruh operators reads
\begin{equation}
\label{eq:nucudef}
\begin{split}
C&\doteq \mathrm{Tr}\left [\hat \rho \left \{ V_\omega \otimes V_{\omega'}^\dagger \right \} \right ] \\
&= \delta(\om - \om') \times \left [ 2 \left (
\begin{array}{cccc}
n_k    &  0  &  0    & c_k \\
0      & n_k & c^*_k &0 \\
0      & c_k & n_k   &0 \\
c^*_k  &  0  &  0    & n_{k}
\end{array} \right )+{1 }\right ]~, 
\end{split}
\end{equation}
where $n_k$ and $c_k$ are given in Eq.~\eqref{eq:nkckdef}.

Using the matrix $B_\om$ of Eq.~\eqref{eq:changeofbase}, and dropping the trivial factor of $\delta(\om - \om') $, the covariance matrix of $R$ and $L$ operators is 
\begin{equation}
\begin{split}
C^{RL}_\omega &= B_\omega\,  C\,  B_\omega^\dagger \\
&= 2 \left (
\begin{array}{cccc}
n_\om    &  c_\om   &  m_\om^*    &  c_\om^{UV} \\
c_\om^*     & n_\om  & c_\om^{UV\,*} &  m_\om   \\
m_\om      & c_\om^{UV} & n_\om    & c_\om   \\
 c_\om^{UV\,*} &  m_\om^* &   c_\om^*   & n_\om 
\end{array} \right )+{1 } ~, 
\end{split}
\label{covariancematrices}
\end{equation}
where
\begin{subequations}
\label{nucu}
\begin{align}
2 n_\omega +1 &= \left ( \abs{\alpha_\omega^{H}}^2 + \abs{\beta_\omega^{H} }^2 \right)  \left (2 n_k+1\right )~, \label{nom} \\
c_\omega&= \alpha_\omega^{H} (\beta_\omega^{H})^* \left (2 n_k+1  \right ) ~, \label{com} \\
m_\omega &=2 {\rm Re}\left( c_k \alpha_\omega^{H} \beta_\omega^{H} \right)~, \\
2 c^{UV}_\om &=   (\alpha_\omega^{H} )^2c_k + \left [ (\beta_\omega^{H} )^{2}c_k\right ]^*~.
\end{align}
\end{subequations}
The first two coefficients concern separately either the $U$, or the $V$-modes. They fix the spectrum and the strength of the correlations. The last two concern the $U-V$ mode mixing, and are proportional to $c_k$.

\begin{figure}[tb]
\centerline{\includegraphics[width=0.95\columnwidth]{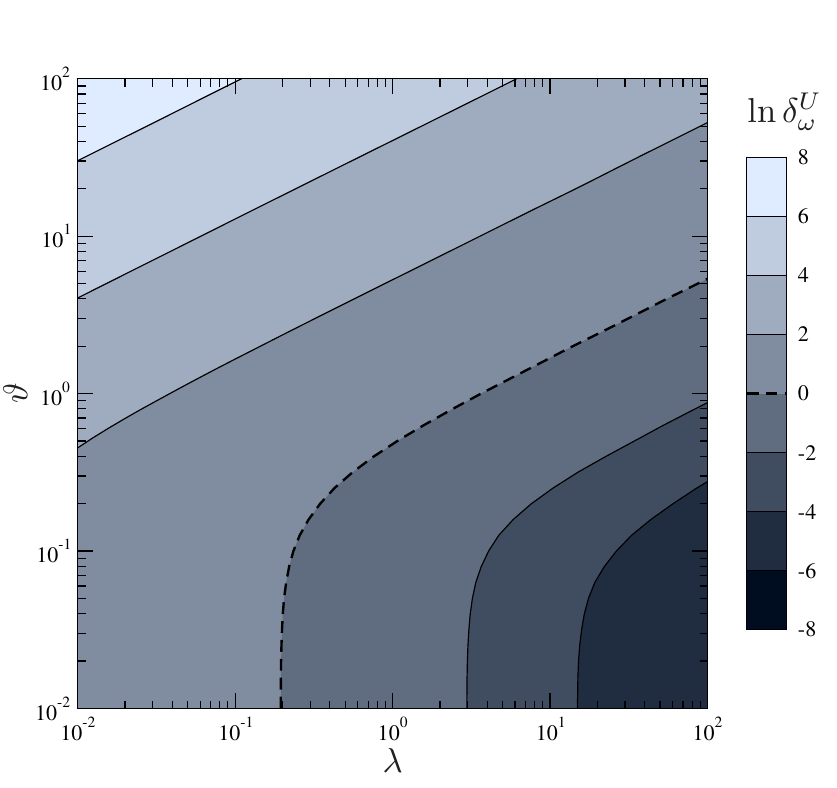}}
\caption{\label{fig:massless-n} \small
Figure for $\delta_\omega^U$ with $\omega = H$ for massless field with critical coupling $g=g_{\rm crit}$. In the infalling vacuum, for $T_\psi = 0$, the nonseparability found for the massless relativistic case is preserved as long as $\Lambda/H = \lambda \gtrsim 1/5$. When the environment is characterized by a temperature $T_\psi \neq 0$, the entanglement is preserved as long as $T_\psi \lesssim  \sqrt{H \Lambda}/ 2$, as explained in the text.}
\end{figure}

Considering the coherence amongst pairs of $U$-quanta, i.e., ignoring the $V$-modes, as in Eq.~\eqref{delta}, we define 
\begin{equation}
\delta_U^\omega \doteq n_\omega +1 - \abs{c_\omega}^2/n_\omega~.
\end{equation}
Using Eq.~\eqref{nucu}, we obtain 
\begin{equation}
\delta_U^\omega = \frac{n_k(n_k+1)}{\left ( |\alpha_\omega^{H}|^2 + |\beta_\omega^{H}|^2 \right) n_k + |\beta_\omega^{H}|^2 }~.
\label{dU}
\end{equation}
We see that $\delta_U$ does not depend on $c_k$. This is to be expected since $c_k$ characterizes the correlation between modes of opposite momenta, and since there is no $U-V$ mode mixing for two-dimensional massless fields. More importantly, \eq{dU} is valid irrespectively of the temperature of the environment $T_\psi$. We can thus study how the separability of $U$-quanta is affected by $T_\psi$. The criterion of nonseparability, $\delta_U^\om < 1$, gives 
\begin{equation}
|\beta_\omega^{H}|^2 = \frac{1}{e^{\om/T_{\rm GH}} - 1}>  \frac{n^2_k(T_\psi)}{2 n_k(T_\psi) +1}~,
\label{dUT}
\end{equation}
where $n_k(T_\psi)$ is plotted in Fig.~\ref{fig:massless-T}. Using this Figure, in Fig.~\ref{fig:massless-n} we study $\ln \delta_\omega^U$ with $\omega = H$ as a function of $\lambda$ and $\vartheta = T_\psi/H$. At zero temperature $T_\psi = 0$, we see that the pair of $U$-quanta with $\om = H$ is nonseparable for $\lambda \gtrsim 0.2$, i.e., for a rather strong dissipation since $\Lambda = H/5$. Using \eq{dUT} we see that this is also true for all quanta with $\om/ H \lesssim 1$. More surprisingly, when $\lambda$ is high enough, this pair is nonseparable {\it even} when $T_\psi > T_{\rm GH}$, i.e., when the environment possesses a temperature higher than the Gibbons-Hawking temperature. Indeed, whenever $T_\psi \lesssim  \sqrt{H \Lambda}/2$, the pair is nonseparable, as all pairs with smaller frequency $\om$.

In other words the quantum entanglement of the lowfrequency $U$ pairs of quanta is extremely {robust} when working with dissipative fields which are relativistic in the infrared. The robustness essentially follows from the kinematical character of the transformation of \eq{eq:changeofbase} which relates two {\it relativistic} mode bases. It is also due to the fact that $n_k$, the number of $U-V$ pairs created by the cosmological expansion, remains negligible in \eq{nucu} as long as $ 1 \ll \Lambda/H $, and $T_\psi \ll  T_{\rm GH} (\Lambda/H)^{1/2} $. 

\section{Black hole radiation}
\label{sec:BHdScorr}

We now explain when and why the above results apply to the Hawking radiation emitted by dissipative fields. We shall be more qualitative than in the former sections because several approximations are involved in the correspondence between de Sitter and the black hole case. Our main aim is to establish that the spectrum of Hawking radiation, and the associated long distance correlations across the horizon, are both robust when dissipation occurs at sufficiently high energy with respect to the surface gravity, as was anticipated in Refs.~\cite{Brout:1995wp,Parentani:2007uq}. 

The robustness shall be established by studying the anticommutator of \eq{eq:GacomXX}, and showing that its asymptotic behavior is governed by Eqs.~\eqref{nom} and~\eqref{com}.

Firstly, being covariant, the action of \eq{eq:covaction} applies as such to any black hole metric endowed with a preferred frame described by a timelike field $u$.\footnote{
When completing this work, we became aware of Refs.~\cite{Lombardo:2012cb,Lombardo:2012xh} where similar issues are discussed. While the model is similar to that of \eq{eq:covaction}, the preferred frame is taken at rest with respect to the orbits of the stationary Killing field $K_t$. This means that $u$ is spacelike in the supersonic region. Unlike what is claimed, we believe that dissipation will necessarily engender an instability. More generally, we have not been able to follow the mathematical developments of these works.}
Secondly, the correspondence with de Sitter becomes more precise when working with stationary settings. At the level of the background, this means that there is a Killing field $K_t$, and that $u$ commutes with $K_t$. In this case, the metric can be written as 
\begin{equation}
\label{eq:PGBH}
ds^2 = - dt^2 + (dX - v(X)dt)^2 ~.
\end{equation}
As in Eq.~\eqref{eq:PG}, $t,X$ are defined by $dt = u^{\mathrm{ff}}_\mu dx^\mu$, and $\partial_X = s_{\mathrm{ff}}^\mu \partial_\mu$, where $u^{\mathrm{ff}}$ is a stationary and freely falling unit timelike field. In the present case, it is no longer unique because the system is no longer translation invariant. It belongs to a one parameter family, where the parameter can be taken to be the value of $v$ at spatial infinity~\cite{Jacobson:2007jx}. When the preferred field $u$ is freely falling (as we shall assume for simplicity), this residual invariance is lifted by working with $u^{\mathrm{ff}} = u$.

By stationary settings, we also meant that the state of the environment is stationary. This implies that the noise kernel of \eq{eq:disskernel} only depends on $t-t'$ when evaluated at $X, X'$, along the orbits of the Killing field $K_t$. When these stationary conditions are met, the (driven part of the) anticommutator of $\phi$ is (exactly) given by \eq{eq:GacomXX}, where the two kernels $G^\om_{\rm ret}$ and $N^\om$ are now defined in the black hole metric of \eq{eq:PGBH}. 

As a result, to compare the expressions of $G^\om_{\rm ac}(X,X')$ evaluated in de Sitter and in \eq{eq:PGBH}, it is sufficient to study $G^\om_{\rm ret}$ and $N^\om$. To establish the correspondence with controlled approximations, the following four conditions are necessary:
\begin{itemize}
\item the state of $\psi$ should be the same
\item the black hole surface gravity $\kappa = H$
\item the near horizon region should be large enough
\item the dispersive and dissipative scales should both be much larger than $\kappa$. 
\end{itemize}
The first condition is rather obvious and needs no justification. The second and the third conditions concern the metric and the field $u$. To characterize the near horizon region (NHR) explicitly, we shall use
\begin{equation}
\begin{split}
v &= - 1 + D \tanh(\kappa X/D) \\
&\sim - 1 + \kappa X + D\,  O(\kappa X/D)^3 ~,
\label{vD}
\end{split}
\end{equation}
which possesses a future (black hole) Killing horizon at $X = 0$. The NHR is defined by the region $|\kappa X |\lesssim D/2$ where $v$ is approximately linear. Hence it is a portion of de Sitter space with $H = \kappa $, see Eq.~\eqref{eq:PG}. It should be emphasized that the mapping also applies to the $u$ field. In fact, when $u$ is freely falling, the only scalar quantity which is involved in the mapping is its expansion evaluated at the horizon: $\Theta_0 = - \nabla_\mu u^\mu =\kappa$. Hence, in the NHR, the orbits of $u$ coincide with those found in de Sitter. (When $u$ is accelerating, both $\Theta_0$ and the acceleration $\gamma_0$ must match, see Eq.~\eqref{gammaindSinapp} and footnote 4 in Ref.~\cite{Busch:2012ne}.) 

Using \eq{vD}, the third condition means that $D$ cannot be too small. This condition was found in Ref.~\cite{Macher:2009tw} when considering the spectral deviations of Hawking radiation which are due to highfrequency dispersion, see also Refs.~\cite{Finazzi:2010yq,Coutant:2011in,Finazzi:2012iu}. For quartic dispersion, these deviations are small when $D^{3/2} \gg \kappa/\Lambda$. In this case, the nontrivial dispersive effects all occur deep inside the NHR, i.e., in a portion of de~Sitter space. Moreover, at fixed $\kappa/\Lambda$, the spectral deviations increase when $D$ decreases. We shall see below that these facts also apply to dissipative fields when the above four conditions are met.

\subsection{The stationary noise kernel}

When considering the model of \eq{eq:covaction} in the metric \eq{eq:PGBH} with $u$ freely falling, the noise kernel $N^\om$ of \eq{eq:GacomXX} is
\begin{align}
\label{Nbh}
N^\om(X_1,X_2) = &\gamma(-\partial_1 )\gamma(-\partial_2 ) \left ( -i\omega + \sqrt{v_1} \partial_1 \sqrt{v_1}   \right )  \\ 
\nonumber & \left ( i\omega + \sqrt{v_2} \partial_2 \sqrt{v_2}   \right )  \int\!\!dq \, G_{\mathrm{ac}, \psi}^\omega(X_1,X_2, q)~,
\end{align} 
where $v_i \doteq v(X_i)$ and $\partial_i \doteq \partial_{X_i}$. The stationary kernel of the last line is the Fourier transform of the anticommutator of $\psi$, see \eq{eq:NofDelta}. To compute it we use the fact that the factor $a(\tau,z)$ of
Eq.~\eqref{eq:Duincov} is now given by (see Eq.~(55) in Ref.~\cite{Parentani:2007uq} for a three-dimensional radial flow) 
\begin{equation}
\begin{split}
a(X,t) = v(X)/v(z(X,t) )~.
\end{split}
\end{equation}
As in Eq.~\eqref{eq:Duincov}, $z$ labels the orbits of $u$. It is here completely fixed by the condition that $z =X$ when $t=0$. Since the orbits are solutions of $dX/dt = v$, $z$ is implicitly given by
\begin{equation}
\begin{split}
\int_{z}^{X} \frac{d X_1}{ v_1} &= t~.
 \end{split}
\end{equation}
Using the above equations to re-express the $\delta(z- z')$ of \eq{eq:NofDelta}, one finds 
\begin{equation}
\begin{split}
G_{\mathrm{ac}, \psi}(\Delta t, X_1,X_2; q) =& \frac{\delta(\Delta t- \int_{X_2}^{X_1} {dX}/{v}) }{\sqrt{v_1 v_2}}  \\
& \times \frac{2n_q+1}{\Omega_q} \cos\left (\Omega_q \Delta t\right ) ~.
\end{split}
\end{equation}
Its Fourier component with respect to $\Delta t$ is trivially 
\begin{equation}
\begin{split}
G_{\mathrm{ac}, \psi}^\omega (X_1,X_2; q) =&   \frac{ \ep{i \omega \Delta t_{12} } }{\sqrt{v_1 v_2}} \ \frac{2n_q+1}{\Omega_q} \cos\left (\Omega_q \Delta t_{12} \right )~, 
\end{split}
\end{equation}
where $\Delta t_{12} = \int_{X_2}^{X_1} {dX}/{v} $ is the lapse of time from $X_2$ to $X_1$ following an orbit $z=cst.$ which connects these two points. Since the settings are stationary, these orbits are all the same, as can be seen in Fig.~\ref{fig:caracter}. 

Using \eq{Nbh}, the noise kernel is explicitly given by
\begin{equation}
\begin{split}
N^\om(X_1,X_2) = &\gamma(-\partial_{1})\gamma(-\partial_{2}) \frac{ \ep{i \omega \Delta t_{12} }}{\sqrt{v_1 v_2}}\\
&\times \int\!\!dq {(2n_q+1)} \, {\Omega_q} \cos\left (\Omega_q \Delta t_{12} \right ) ~.
\label{Nbh2}
\end{split}
\end{equation}
This kernel is local in that it only depends on $\mathsf{g}_{\mu \nu}$ and $u^\mu$ between $X_1$ and $X_2$. Hence, when evaluated in the black hole NHR, it agrees, {\it as an identity}, with the corresponding expression evaluated in de Sitter.

In conclusion, we notice that this identity follows from our choice of the action of \eq{eq:covaction}. Had we used a more complicated environment, this identity would have been replaced by an approximative correspondence. In that case, the correspondence would have still been accurate if the propagation of $\psi$ had been {adiabatic}. As usual, this condition is satisfied when the degrees of freedom of $\psi$ are \enquote{heavy}, i.e., when their frequency $\Omega_q \sim \Lambda \gg \kappa$. 

\subsection{The stationary \texorpdfstring{$G_{\rm ret}^\om$}{Gret(omega)}}

\begin{figure}
\centerline{\includegraphics[width=1\columnwidth]{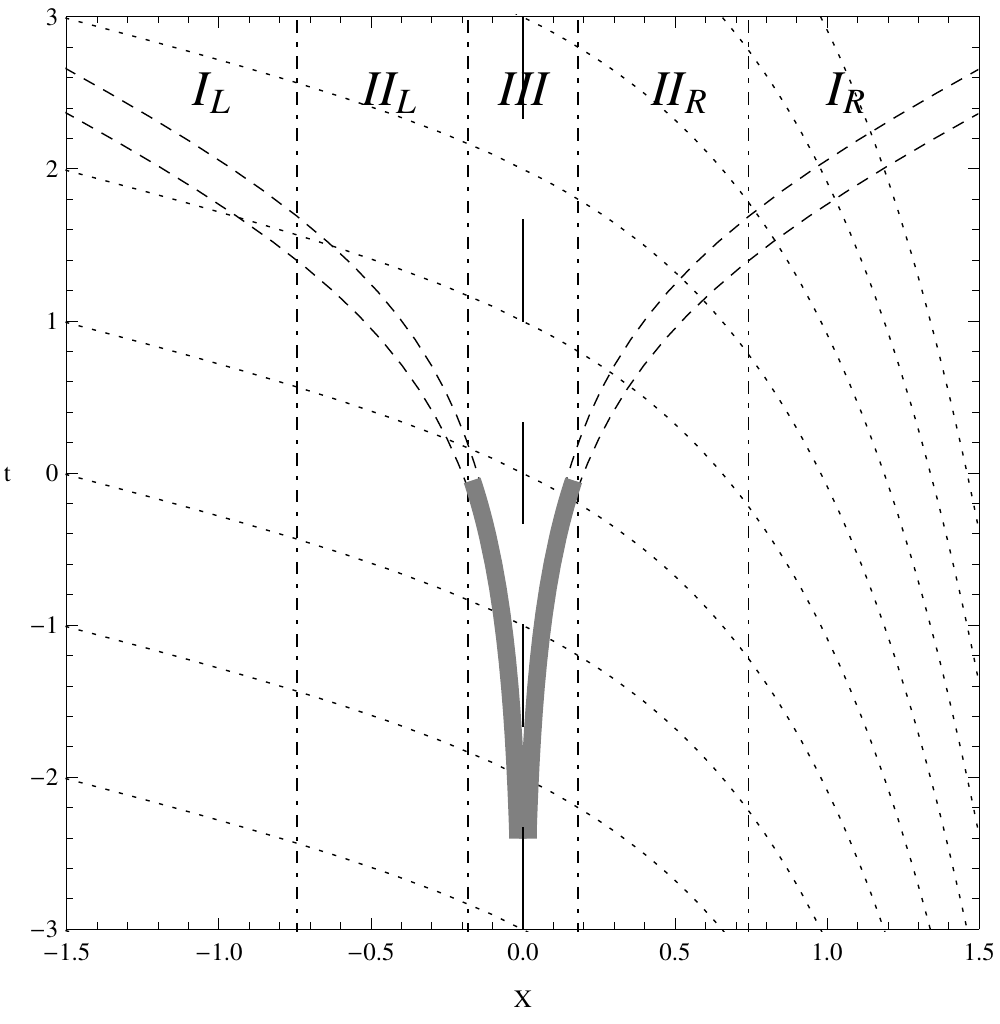}}
\caption{\label{fig:caracter} \small
Null outgoing geodesics (dashed lines) on either side of the horizon at $X=0$, and freely falling orbits $z=\mathit{cst}.$ (dotted) in the $t,X$ coordinates of \eq{eq:PGBH}. As explained in the text, the nearby geodesics schematically indicate the space-time region where $G_{\rm ac}(t,X,t_1,X_1)$ is nonvanishing, when $\kappa t_1=2.5$, and $\kappa X_1= 1.5$, see Fig.~1 of Ref.~\cite{Parentani:2010bn} for the relativistic case. The two thick solid lines represent the region where the noise kernel contributes to $G_{\rm ac}$, see \eq{eq:GacomXX} and \eq{eq:GacXXP}. In the central region \textit{III}, the propagation is well described in $P$ space, and resembles to that found in de~Sitter.}
\end{figure}

The stationary function $G_{\rm ret}^\om(X,X_1)$ obeys Eq.~\eqref{locEq}, which is a fourth order equation in $\partial_X$ when working with Eq.~\eqref{eq:numdispersion}. Depending on the position of $X$ and $X_1$, its behavior should be analyzed using different techniques. Far away from the horizon, the propagation is well described by WKB techniques since the gradient of $v$ is small. Close to the horizon instead, the WKB approximation fails, as in dispersive theories~\cite{Coutant:2011in}. In this region, the $P$ representation accurately describes the field propagation, and is essentially the same as that taking place in de Sitter. Therefore, the calculation of $G_{\rm ac}^\om(X,X')$ of \eq{eq:GacomXX} at large distances boils down to connecting the de Sitter--like outcome at high $P$ to the low-momentum WKB modes. As in the case of dispersive fields, the connection entails an inverse Fourier transform from $P$ to $X$ space in the intermediate region \textit{II}, see Fig.~\ref{fig:caracter}, where both descriptions are valid~\cite{Brout:1995wp,Corley:1997pr,Unruh:2004zk,Balbinot:2006ua,Coutant:2011in}. In the present case, these steps are performed at the level of the two-point function rather than being applied to stationary modes. In fact, we shall compute $G_{\rm ac}^\om$ through 
\begin{equation}
\begin{split}
\label{eq:GacXXP}
G_{\rm ac}^\omega(X,X')\!  &= \!  \int\!\!\! \! \int^\infty_{-\infty}\! \, d\bP_1 d\bP_2 G_{\rm ret}^\omega(X,\bP_1) \\
 &\hspace{1cm} \times G_{\rm ret}^{\omega\,*}(X',\bP_2) N^\omega(\bP_1,\bP_2) ~,
\end{split}
\end{equation}
where the two $G_{\rm ret}^\om$ are expressed in a mixed $X,P$ representation. The early configurations in interaction with the environment are described in $P$ space, while the large distance behavior is expressed in $X$ space. 

Let us give here only the essential points, more details are given in Appendix~\ref{app:BHdS}. The validity of the whole procedure relies on a combination of the third and the fourth condition given above, namely ${\rm max}(1, D^{-2}) \ll \Lambda/\kappa $, and is limited to moderate frequencies, i.e., $ 0 < \omega \sim \kappa \ll \Lambda$. 

For simplicity, we consider massless fields. Then $\Lambda/\kappa \gg 1$ guarantees that the infalling $V$ modes essentially decouple from the outgoing $U$ modes because the only source of $U-V$ mixing comes from the ultraviolet sector. Hence, at leading order in $\kappa/\Lambda$, it is legitimate to consider only the $U$ modes. For massive fields with $m \ll \Lambda$, the discussion is more elaborate but the main conclusion is the same: the properties of the Hawking radiation are robust.

For massless fields, at fixed $\omega$, the propagation of the $U$ modes is governed by the effective dispersion relation, see Eq.~\eqref{eq:eom-transformed},
\begin{equation}
\label{HJeqU}
\Omega = \om - v(X) P = \sqrt{F^2- \Gamma^2} ~.
\end{equation} 
As long as $P \ll \Lambda$, the $U$ sector of $G_{\rm ret}^\om$ behaves as for a relativistic field, since $\sqrt{F^2 - \Gamma^2}\sim P (1 + O(P/\Lambda))$. Instead, when $P \gtrsim \Lambda$, the dispersive and dissipative terms weighted by $f$ and $\Gamma$ cannot be neglected in Eq.~\eqref{locEq}. To characterize the transition from these two regimes, we consider the optical depth of \eq{eq:opticaldepth}. When working at fixed $\om$, one finds 
\begin{equation}
\label{eq:optdepthBH}
\begin{split}
 \mathcal{I}_\om(X,X_1) &= \int_{P}^{P_1} dP' \frac{\Gamma(P')}{ P' \, \partial_X v\left [ X_\omega(P') \right ]}~, \\
 &= \int^{X}_{X_1} d X'  \frac{\Gamma[P_\omega(X')]}{v^\om_{\rm gr}(X') }~,
\end{split}
\end{equation}
where $X_\omega(P)$ is the root of~\eq{HJeqU}, as is $P_\om(X)$ when using $X$ as the variable. The first expression governs $G_{\rm ret}$ in the NHR where $\partial_X v \sim \kappa$ is almost constant, see \eq{eq:Gret-solution}. To leading order in $\Gamma/P \ll 1$, which is satisfied everywhere but very close to the horizon, the second expression governs $G_{\rm ret}$ in $X$ space. Since $v_{\rm gr}^\om= 1/\partial_\om P $ is the group velocity in the rest frame, $\mathcal{I}_\om  = \int^t_{t_1} dt' \Gamma(P_\om)$, where the integral is evaluated along the classical outgoing trajectory. It should be noticed that, when considered in $X$ space, $ \mathcal{I}_\om$ applies on the right and the left of the horizon. In the R region, $v_{\rm gr} > 0$, while it is negative in L, so that in both cases $\mathcal{I}_\om > 0$ when $P_1 > P> 0$, i.e., when $P_1$ is in the past of $P$. 

To characterize the retarded Green functions of \eq{eq:GacXXP}, we compute $\mathcal{I}_\om$ in the mixed representation, in the limit where $P_1$ is large enough so that $X_\om(P_1)$ is deep inside the NHR, while $X$ is far away from that region. For simplicity, we consider the case of Eq.~\eqref{eq:numdispersion} with $g=g_{\rm crit}$. In this case, only the dissipative effects are significant,\footnote{
In the case where $g^2 \ll 1$, dispersive effects are important and may limit the role of dissipation in the NHR. In that case, the decaying part of the field will contribute to Eq.~\eqref{eq:Gacdef}. This situation corresponds to what is found in the surface wave experiments~\cite{Weinfurtner:2010nu,Rousseaux:2007is}.} 
and one finds
\begin{equation}
\label{eq:optdepthBHlimit}
\begin{split}
 \mathcal{I}_\om(X,P_1) & \sim  \frac{P_1^2}{2\kappa \Lambda}  + \frac{\omega^2 |X| }{\Lambda |1+v_{R/L}|^3}~, 
\end{split}
\end{equation}
where $v_R$ ($v_L$) is the asymptotic velocity on the right (left) side.
From the second term, we learn that $|\kappa X|$ should be much smaller than $\Lambda/\kappa$ for the Hawking quanta not to be dissipated. Since we work in the regime $\Lambda/\kappa \gg 1$, this condition is easily satisfied. We notice that a similar type of weak damping effect of outgoing modes has been observed in experiments~\cite{Weinfurtner:2010nu}.

\begin{figure}
\centerline{\includegraphics[width=1\columnwidth, height=6cm]{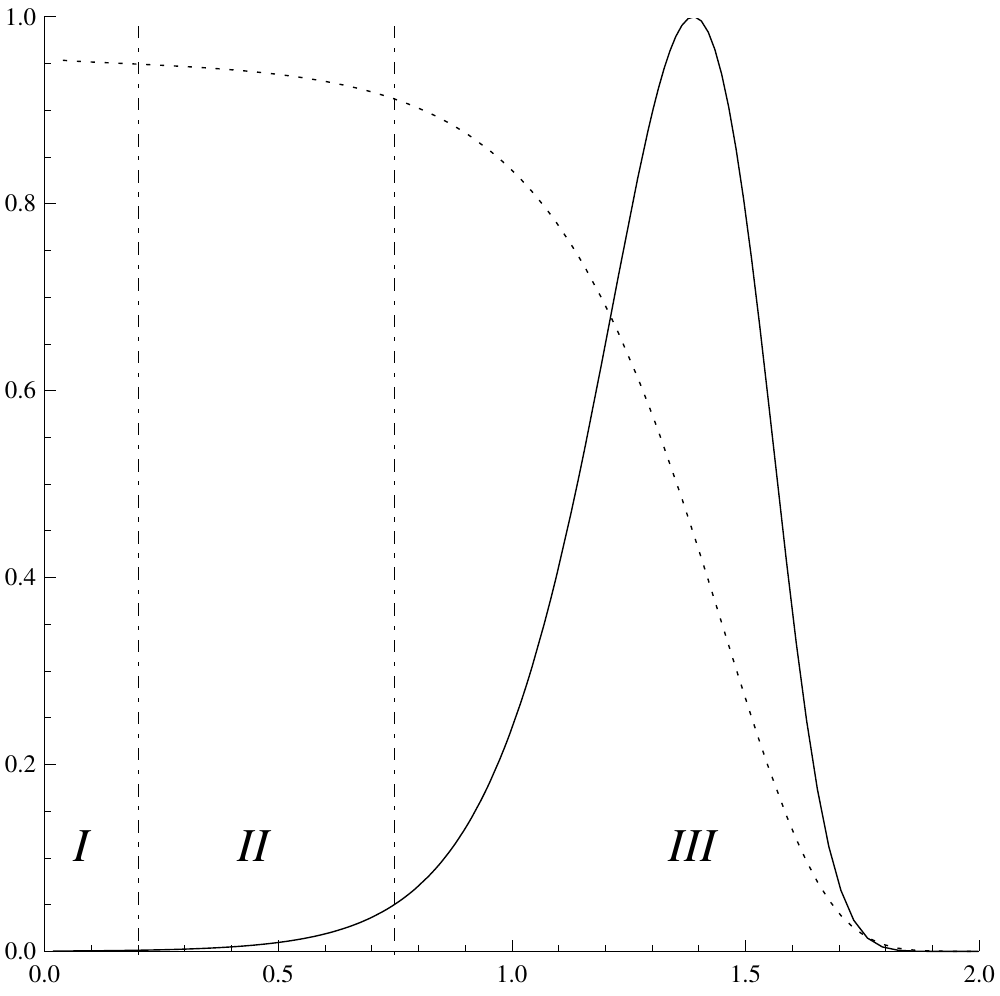}}
\caption{\label{fig:velocity}
As a function of $\log_{10} (P/\kappa)$, in a dotted line we plot $\exp\{- \mathcal{I_\om}(X,P)\}$, the optical depth of Eq.~\eqref{eq:optdepthBH}, evaluated for $\om/\kappa = 1$, $\Lambda /\kappa = 400$, and $\kappa X = 20$. The solid curve represents $T(P)$ of Eq.~\eqref{eq:dominPT} for the same values, and $D = 0.99$. The left dash-dotted line corresponds to the limit of the NHR: $\kappa X = D/2$, here reexpressed as $P = 2\kappa / D $. For lower $P$, in region \textit{I}, the de Sitter--like $P$ representation fails. The right vertical line indicates the upper limit of the $X$-WKB approximation, see Appendix \ref{app:BHdS}. For the adopted values, the region \textit{II} where the $P$ and the $X$ descriptions are both valid has a finite size. We also see that $T$ vanishes in region \textit{I}.}
\end{figure}

From the first term, we learn that $\mathcal{I}_\om$ gives an upper bound to the domain of $P$ which significantly contributes to \eq{eq:GacXXP}, namely $P^2 \lesssim \Lambda \kappa$, as in de Sitter. A lower bound of this domain is provided by the $\gamma$ factors of \eq{eq:noisekernel}. Using this equation and \eq{prepomega}, the integrand of \eq{eq:GacXXP} scales as 
\begin{equation}
\label{eq:dominPT}
\begin{split}
T(P) \propto & P \, \Gamma(P) \ep{- 2 \mathcal{I} ( X , P ) } \propto P^3 e^{-P^2/\Lambda \kappa}~,
\end{split}
\end{equation}
and its behavior is represented in Fig.~\ref{fig:velocity}. Hence, the relevant domain of $P$, i.e., when $T$ is larger than $10\%$ of its maximum value, scales as 
\begin{equation}
\label{eq:dominP}
\begin{split}
0.36 \sqrt{\kappa \Lambda} = P_{\rm min} \lesssim  P \lesssim  P_{\rm max} = 2.4 \sqrt{\kappa \Lambda}~. 
\end{split}
\end{equation}
Considered in space-time, since $P \sim e^{-\kappa t}$, this limits the lapse of time during which the coupling to $\psi$ occurs. Interestingly, this lapse is given by $\kappa \Delta t \approx 2$, i.e., two e-folds, irrespective of the value of $\Lambda/\kappa$, and that of $\om$. It should be also stressed that nothing precise can be said about the domain of $X$, which significantly contributes because the $X$-WKB fails when $P$ is so large. One can simply say that it is roughly characterized by the interval $[- X_{\rm trans},  X_{\rm trans}]$, where $X_{\rm trans} = X_{\om = \kappa}(P_{\rm min})$ is given by 
\begin{equation}
\label{Xtrans} 
\kappa X_{\rm trans} \sim 3 \sqrt{\kappa / \Lambda}~. 
\end{equation} 
This value defines the central region \textit{III}, see Fig.~\ref{fig:velocity} and Fig.~\ref{fig:caracter}. Using the profile of \eq{vD}, $ X_{\rm trans}$ is situated deep inside the NHR when $\kappa / \om_{\rm max}^{\rm diss} \ll 1$, where the critical frequency $\om_{\rm max}^{\rm diss}$ is given by
\begin{equation}
\label{eq:ommdiss}
\begin{split}
\om_{\rm max}^{\rm diss} = \Lambda D^2 ~.
\end{split}
\end{equation}

Hence, when $\kappa / \om_{\rm max}^{\rm diss} \ll 1$, the coupling between $\phi$ and $\psi$ is accurately described in the $P$ representation, and takes place in a portion of de Sitter. In addition, the connection between the high- and low-momentum propagation can be safely done in the intermediate region \textit{II}, defined by $\kappa |X_{\rm trans}| \ll \kappa| X | \lesssim  D $, see Fig.~\ref{fig:caracter}, where, on the one hand, one is still in a de Sitter--like space since $v$ is still linear in $X$, and, on the other hand, the low-momentum modes can be already well approximated by their WKB expressions. Notice finally that this reasoning only applies for frequencies $\om  \ll \om_{\rm max}^{\rm diss} $. Indeed, when $\om = \om_{\rm max}^{\rm diss}$, dissipation occurs around $\kappa X \sim D$, i.e. no longer in a de Sitter like background. 

These steps are sufficient to establish that the results of Sec.~\ref{Asc} apply for $\om \ll \om_{\rm max}^{\rm diss}$. In particular, \eq{nom} implies that the spectrum of radiation is robust (when the temperature of the environment is low enough, see Fig.~\ref{fig:massless-n}). Namely, to leading order in $\kappa/\Lambda$, the mean occupation number $n_\om$ of quanta received far away is given by the Planck distribution at the standard relativistic temperature $T_\mathrm{H} = \kappa/2\pi$. As in dispersive settings, the real difficulty is to evaluate the spectral deviations. In this respect, we conjecture that the leading deviations due to dissipation will be suppressed by powers of $\kappa/\om_{\rm max}^{\rm diss}$. That is, they will be governed by the composite ultraviolet scale of \eq{eq:ommdiss} which depends on the high-energy physics, here with $\Gamma$ quadratic in $P$, and on the extension $D$ of the black hole NHR. This second dependence is highly relevant when $D \ll 1$. 

Together with the robustness of the spectrum, one also has that of the long-distance correlations across the horizon between the Hawking quanta and their partners. These correlations are fixed by the coefficient $c_\om$ of \eq{com}. To get the space-time properties of the pattern, one should integrate over $\om$, i.e., perform the inverse Fourier transform of Eq.~\eqref{eq:fouriert}, because it is this integral that introduces the space-time coherence~\cite{Massar:1995im,Brout:1995wp,Parentani:2010bn}. In Fig.~\ref{fig:caracter}, we have schematically represented the anticommutator $G_{\rm ac}(t-t_1,X,X_1)$ in the $t-t_1,X$ plane when $X_1$ is taken far away from the horizon.

\begin{itemize}

\item
Far away from the NHR, in regions $I_R$ and $I_L$, for $D \lesssim |\kappa X| $, the characteristics of the field follow null geodesics, see \eq{retism}. Since $v \sim cst.$, they no longer separate from each other. Hence, at large distances, the space-time pattern obtained by fixing one point~\cite{Massar:1995im,*Massar:1996tx}, and the equal time correlation pattern~\cite{Balbinot:2007de}, will be the same as those predicted by a relativistic treatment. 

\item
In the two intermediate regions \textit{II}$_R$ and \textit{II}$_L$, for $ X_{\rm trans} \lesssim |\kappa X| \lesssim D $, the characteristics separate from each other following $\delta X \sim e^{\kappa t}$ since their behavior is already close to the relativistic one. This pattern is obtained by considering two-point functions with one point fixed, or wavepackets~\cite{Brout:1995wp}. It is interesting to notice that it cannot be obtained by considering equal time correlations, since these develop only outside the NHR, for $|\kappa X| \gtrsim D$~\cite{Parentani:2010bn}. Indeed as long as $X$ and $X'$ are in the NHR, the (approximate) de Sitter invariance under $K_z$, see Appendix \ref{App:Prep}, implies that $\int d\om G^\om_{ac}$ only depends on $X-X'$.~\footnote{From this observation, we learn that the correspondence between the physics in black hole metrics and in de Sitter is not merely a convenient way to obtain $n_\om$ and $c_\om$ in Appendix \ref{sec:BHds:ds}. It actually shows up in the NHR when computing observables, such as the mean value or the two-point correlation of $\rho = u^\mu u^\nu T_{\mu \nu}$. Moreover, it ceases when leaving this region. In this sense, the Hawking effect only develops, or separates, from its de Sitter roots for $|\kappa X| \gtrsim D$, and furthermore, this separation is adiabatic.}
 
\item
The central region \textit{III} is the region where the configurations of the $\phi$ field are driven by the noise kernel. In Fig.~\ref{fig:caracter} the two thick solid lines indicate the space-time locus where the interactions involving the configurations selected by $t_1, X_1$ are taking place.~\footnote{
To get Fig.~\ref{fig:caracter}, we have filtered out low frequencies. This amounts to considering a wavepacket rather than the two-point function, see Eqs.~(38,39) in Ref.~\cite{Parentani:2010bn}. Had we considered $G_{\rm ac}$, the thick lines would have extended back in time, because the coupling between $\psi$ and $\hat\phi_\om$ is centered around
$\kappa t_\om = \ln \om + cst.$, which fixes the blueshift for $P_\om \propto \om$ to reach $\sqrt{\kappa \Lambda}$, see \eq{eq:dominP}.} 
In this central region \textit{III}, the propagation is well described in $P$ space, and corresponds to that found in de Sitter, see \eq{eq:GcXPchiBH} and \eq{eq:ncac}.

\end{itemize}

In brief, when $\kappa/ \om_{\rm max}^{\rm diss} \ll 1$ and $\om/ \om_{\rm max}^{\rm diss} \ll 1$, the nontrivial propagation only occurs deep inside the NHR which is a portion of de~Sitter space. This implies that $n_\om$ and $c_\om$ are, to a good approximation, given by their de~Sitter expressions of Eq.~\eqref{nucu}. Given that these (exact) expressions hardly differ from the relativistic ones when $\kappa/\Lambda \ll 1$, we can predict that, when computed in a black hole metric, these two observables are robust whenever the finiteness of the NHR introduces small deviations with respect to the de Sitter case. For $\om/ \om_{\rm max}^{\rm diss} \ll 1$, this is guaranteed by $\kappa/ \om_{\rm max}^{\rm diss} \ll 1$.

\section{Conclusions}
\label{sec:conclusions}

In this paper we used (a two-dimensional reduction of) the dissipative model of Ref.~\cite{Parentani:2007uq} to compute the spectral properties and the correlations of pairs produced in an expanding de Sitter space. The terms encoding dissipation in \eq{eq:covaction} break the (local) Lorentz invariance in the ultraviolet sector. Yet, they are introduced in a covariant manner by using a unit timelike vector field $u$ which specifies the preferred frame. In addition, the unitarity of the theory is preserved by coupling the radiation field $\phi$ to an environmental field $\psi$ composed of a dense set of degrees of freedom taken, for simplicity, at rest with respect to the $u$ field. Again for simplicity, the action is quadratic in $\phi,\psi$, and the spectral density of $\psi$ modes is such that the (exact) retarded Green function of $\phi$ obeys a local differential equation, see \eq{eq:boxdiss} and \eq{locEq}.
 
By exploiting the homogeneous character of the settings, we expressed the final occupation number $n_k$, and the pair-correlation amplitude $c_k$, in terms of the noise kernel $N$ and the retarded Green function, see \eq{eq:n_k,c_k}. Rather than working with integrals over time as usually done, we used the proper momentum $P = k /a(t)$ to parametrize the evolution of field configurations. Hence, \eq{eq:n_k,c_k} can be viewed as flow equations in physical momentum space. This possibility is specific to the residual symmetry group found in de Sitter space when the $u$ field commutes with the two Killing fields $K_t$ and $K_z$. These group theoretical aspects are explained in Appendix~\ref{App:Prep}. The key equations are \eq{eq:Prepresentation} and \eq{prepomega} which show how the $P$representation is related to the invariant distances, to the homogeneous representation of \eq{Prep2ptfn}, and to the stationary one. This representation is extended to Feynman rules and Schwinger-Dyson equations of (relativistic) interacting field theories in Ref.~\cite{Parentani:2012tx}. 

We numerically computed $n_k$ and $c_k$ in Sec.~\ref{sec:homogen}. When considering a massless field, $n_k$ and the strength of the correlations are plotted as functions of the scale separation $\Lambda/H$, and the temperature of the environment $T_\psi/H$, in Fig.~\ref{fig:massless-T}. The robustness of the relativistic results is established in the limit of a large ratio $\Lambda/H$. The key result concerns the threshold values of the parameters, see the locus $\delta_k = 1$ on the right panel, for which the final state remains nonseparable, i.e., so entangled that it cannot be described by a stochastic ensemble. Various criteria of nonclassicality are compared in Appendix \ref{app:CSclass}. This analysis was then extended to massive fields, see Fig.~\ref{fig:massive-T}, and to the consequences of varying the relative importance of dissipative and dispersive effects, see Fig.~\ref{fig:massive-vac}. As expected, the quantum coherence is lost at high coupling, and when the temperature of the environment is high enough. 

In Sec.~\ref{sec:statio} we exploited the stationarity, and we studied how the thermal distribution characterizing the Gibbons-Hawking effect is affected by dissipation. As in the case of dispersion~\cite{Busch:2012ne}, we found that the thermal character is, to leading order, robust. We also computed the deviations of the effective temperature with respect to the standard one $T_\mathrm{GH} = H/2\pi$, see Fig.~\ref{fig:Tofom} and Fig.~\ref{fig:logToflambda}. In preparation for the analysis of the Hawking effect, we studied the strength of the asymptotic correlations across the Killing horizon between (right) moving quanta with opposite frequency. Quite remarkably, we found that the pairs remain entangled (the two-mode state remains nonseparable) even for an environment temperature exceeding $T_\mathrm{GH} = H/2\pi$, see Fig~\ref{fig:massless-n}. 

Finally, in Sec.~\ref{sec:BHdScorr} we extended our analysis to black hole metrics. When four conditions are met, we showed that the above analysis performed in de~Sitter applies to Hawking radiation. The inequality which ensures the validity of this correspondence is $\kappa/ \om_{\rm max}^{\rm diss} \ll 1 $, where $\om_{\rm max}^{\rm diss}$ is the composite ultraviolet scale of \eq{eq:ommdiss}. It depends on both the microscopic scale $\Lambda$, and $D$, which fixes the extension of the black hole near horizon region where the metric and the field $u$ can be mapped into de~Sitter. The validity of the correspondence in turn guarantees that, to leading order, the Hawking predictions are robust -- even if the early propagation completely differs from the relativistic one, see Fig.~\ref{fig:caracter}. This establishes that when leaving the very high momentum $P \sim \Lambda$ (trans-Planckian) region and starting to propagate freely, the outgoing configurations are ``born'' in their Unruh vacuum state~\cite{Jacobson:1991gr,Jacobson:1993hn,Brout:1995rd}. The microscopic implementation of this state in dissipative theories is shown in \eq{Unv}. As a result, as in the case of dispersive theories~\cite{Macher:2009tw,Coutant:2011in}, the leading deviations with respect to the relativistic expressions should be suppressed as powers of $\kappa/ \om_{\rm max}^{\rm diss}$, i.e., they should be governed by the extension of the black hole NHR which is a portion of de~Sitter space.

In conclusion, even though our results have been derived in $1+1$ dimensions, we believe that very similar results hold in four dimensions, at least for homogeneous cosmological metrics and for spherically symmetric ones, because a change of the dimensionality only affects the low-momentum mode propagation. Hence even if this introduces nontrivial modifications, as grey body factors in black hole metrics, they will not interfere with the high-momentum dissipative effects when the hierarchy of scales $\Lambda/H , \Lambda/\kappa\gg 1$ is found. They can thus be computed separately. 

\acknowledgments

This work has been supported by the FQXi Grant \textit{``Hawking radiation in dissipative field theories''} (No. FQXi-MGB-1129). J.A. wants to thank the Laboratoire de Physique Th\'eorique at Orsay for hospitality and the German Research Foundation (DFG) for financial support through the Research Training Group 1147 ``Theoretical Astrophysics and Particle Physics'' at the University of W\"urzburg, where parts of this work have been carried out. We are grateful to Ted Jacobson and Iacopo Carusotto for interesting remarks.

\appendix
\section{Affine group and P representation}
\label{App:Prep}

We remind the reader that the affine group is the subgroup of the de Sitter isometry group which is generated by the Killing fields $ K_z = \partial_z\vert_t$ and $K_t = \partial_t \vert_X$, which possess the following commutator $ [ K_z , K_t] = - H K_z$. The definition of the coordinates $t,z,X$ is given in \eq{eq:FRW} and \eq{eq:PG}. In de Sitter space, there are two geometrical invariants under this group. Using the coordinates $t,X$, they read 
\begin{subequations}
\label{eq:invariantparameters}
\begin{align}
\label{eq:firstivariant}
\Delta_1 &= \ep{H(t-t')} ~,\\
\Delta_2 &= X\ep{-H(t-t')/2} -X'\ep{H(t-t')/2}~. 
\label{eq:secondinvariant}
\end{align}
\end{subequations}
They are linked to the de Sitter invariant distance by 
\begin{equation}
\label{eq:distdesitter}
\Delta^2 = \Delta_2^2 - (\Delta_1-\frac{1}{\Delta_1})^2 ~. 
\end{equation}
The distances $\Delta_1, \Delta_2$ can also be defined in a coordinate invariant manner. The interested reader will find the expressions at the end of this Appendix. 

When working with states that are invariant under the affine group, the n-point correlation functions only depend on $\Delta_1$ and $\Delta_2$ evaluated between the various pairs of points. Hence, any two-point functions $G_\mathrm{any}(\mathsf x,\mathsf x')$ can be written as $ \tilde G_\mathrm{any} \left (\Delta_1(\mathsf x,\mathsf x'),\Delta_2(\mathsf x,\mathsf x') \right )$. However, it turns out that it is {\it not} convenient to use $\Delta_1, \, \Delta_2$ to compute \eq{eq:acgrennfunction}, and this even though the four integrals of that equation can be easily expressed in terms of two over $\Delta_1$ and two over $\Delta_2$. The reason is that the integrals over the $\Delta_2$ are convolutions. Hence, it is appropriate to work with the Fourier transform with respect to $\Delta_2$ because, in this representation, \eq{eq:acgrennfunction} contains only two integrals. 

The fact that only two variables are needed is not a surprise, given the homogeneity (stationarity) of the setting. Indeed using $G^k(t,t')$ ($G^\omega(X,X')$) of \eq{eq:fourierkt}, one immediately has
\begin{subequations}
\begin{align}
\label{eq:Gacktt}
G_{\rm ac}^k(t,t')\! \! &=\!\!\int\!\!\int\! \!  dt_1 dt_2 G_{\rm ret}^k(t,t_1) G_{\rm ret}^{k\,*}(t',t_2) N^k(t_1,t_2)~,\\
\label{eq:GacomXX}
G_{\rm ac}^\omega(X,X')\! \! &=\!\!\int\! \!  \int\! \!  dX_1 dX_2 G_{\rm ret}^\omega(X,X_1) \\
\nonumber &\hspace{1cm} \times G_{\rm ret}^{\omega\,*}(X',X_2) N^\omega(X_1,X_2) ~.
\end{align}
\end{subequations}
To understand the relationship between these two re\-presen\-ta\-tions, it turns out that the most convenient variables are the proper momenta $P = | s_\mathrm{ff}^\mu p_\mu| $ and $P' = | s_\mathrm{ff}^\mu p'_\mu|$. The reasons for this are many. Firstly, $P$ is invariantly defined; secondly, $\Delta_1$ is easily expressed in $P,P'$ space; thirdly, so is the variable conjugated to $\Delta_2$; and fourthly, $P$ can be attributed to the field itself, so that one can easily take the even (anticommutator) and the odd part of the two-point functions. Let us explain these reasons. 

Once the de Sitter group is broken in a way which preserves the affine group, $P$ is invariantly defined as the momentum associated with the orthogonal fields $u_{\mathrm{ff}},s_{\mathrm{ff}}$ which commute with $K_t$ and $K_z$, and where $u_{\mathrm{ff}}$ is geodesic. In our case, we work with the preferred field $u = u_{\mathrm{ff}}$, but this needs not be the case for $P$ to be unambiguously defined as $P^2 = (s^\mu_{\mathrm{ff}} p_\mu)^2$.

Since $P = k e^{-Ht}$, $\Delta_1$ is simply
\begin{equation}
\label{eq:Delta1}
\Delta_1(\mathsf x,\mathsf x') = P'/P > 0 ~. 
\end{equation}
In addition, the momentum conjugated to $\Delta_2$, defined by $\bar \bP \doteq \partial_{\Delta_2 \vert \Delta_1} $, is given by the geometrical mean
\begin{equation}
\bar \bP =  \bP \sqrt{\Delta_1} = \mathrm{sgn}(\bP) \sqrt{P P'}~.
\end{equation}
The first equality follows from $ \Delta_2 \sqrt{\Delta_1} = X + f(t,t',X')$, and $\bP \doteq \partial_{X \vert t,t',X'}$. The second one follows from \eq{eq:Delta1}. Hence, the Fourier transform of $ \tilde G_\mathrm{any} \left (\Delta_1,\Delta_2 \right )$ with respect to $\Delta_2$,
\begin{equation}
\begin{split}
\label{eq:Prepresentation}
G_\mathrm{any}(\bP,\bP') &=\theta(\bP \bP')   \frac{\sqrt{P P'}}{H} \\
& \int  d\Delta_2\ep{-i \sqrt{P P'} \rm{sgn}(\bP) \Delta_2} \, \tilde G_\mathrm{any} \left (\frac{P'}{P},\Delta_2 \right )~,
\end{split}
\end{equation}
only depends on $\bP$ and $\bP'$. Moreover, if one imposes the isotropy of the setting, $\tilde G_\mathrm{any} \left (\Delta_1,\Delta_2 \right )$ is even in $\Delta_2$, and $G_\mathrm{any}(-\bP,-\bP') = G_\mathrm{any}(\bP,\bP')$. Hence, in this case, all the information is contained in $G_\mathrm{any}(P,P')$. 

The important point is that $G_\mathrm{any}(P,P')$ defined by \eq{eq:Prepresentation} coincides with the lhs of \eq{Prep2ptfn}. In addition, starting with the stationary representation of \eq{eq:fouriert}, one can also verify that the double Fourier transform
\begin{equation}
\label{eq:fourieromP}
G_{\mathrm{any}}^{\omega}(\bP,\bP') = \int \frac{dX dX'}{2\pi } \ep{-i \bP X + i \bP' X'} G_{\mathrm{any}}^{\omega}(X,X')
\end{equation}
has automatically the following structure
\begin{equation}
\begin{split}
\label{prepomega}
G_{\mathrm{any}}^{\omega}(\bP,\bP') &= \frac{(P/P')^{-i \omega/H}}{P P' }G_\mathrm{any} (\bP,\bP')~, 
\end{split}
\end{equation}
where $G_\mathrm{any} (\bP,\bP')$ is given by \eq{eq:Prepresentation}. Together with \eq{Prep2ptfn}, 
\eq{eq:Prepresentation} and \eq{prepomega} are the key equations of this appendix: Whenever a two-point function $G_\mathrm{any} (\mathsf x,\mathsf x')$ is invariant under the affine group, its Fourier transforms $G_\mathrm{any}^k (t,t')$ and $G_\mathrm{any}^\omega (\bP,\bP')$ are related to $G_\mathrm{any} (\bP,\bP')$ of \eq{eq:Prepresentation} by \eq{Prep2ptfn} and \eq{prepomega} respectively. Finally, the antisymmetry of the commutator $G_\mathrm{c}$ is expressed as $G_\mathrm{c}(P',P) = - G_\mathrm{c} (P,P')^* $ while the symmetry of $G_{\rm ac}$ gives $G_{\rm ac}(P',P) =G_{\rm ac} (P,P')^* $.

To conclude this Appendix, we express $\Delta_1$ and $\Delta_2$ in covariant terms. The log of $\Delta_1$ is given by the line integral of $u^{\mathrm{ff}}$ from $\mathsf x$ to $\mathsf x'$, that is
\begin{equation}
\ln \Delta_1= -  H \int_{\mathsf x}^{\mathsf x'}  u^{\mathrm{ff}}_{\mu}\,  dx^\mu ~.
\end{equation}
This is an invariant expression. Indeed, on the one hand, since $u^{\mathrm{ff}}$ is geodesic, $u^{\mathrm{ff}}_{\mu} dx^\mu$ is an exact 1-form and the above integral does not depend on the path. On the other hand, $u^{\mathrm{ff}}$ is the only (timelike) unit geodesic field that commutes with $K_z$ and $K_t$. Since, Eq.~\eqref{eq:distdesitter} gives $\Delta_2$ as a combination of $\Delta_1$ and $\Delta$ which are both invariantly defined, so is $\Delta_2$.~\footnote{If one wishes, $\Delta_2$ can also be seen as the integral of $s^{\mathrm{ff}}_{\mu} dx^\mu $, the 1-form associated to the vector orthogonal to $u^{\mathrm{ff}}$. Since this form is not exact, one has to specify the contour from $\mathsf x$ to $\mathsf x'$. Using the $t,z$ coordinates, one should go at fixed $z$ from $t$ to $(t+t')/2$, then vary $z$ at fixed time until $((t+t')/2,z')$, and vary $t$ at fixed $z$ until $(t',z')$. Any different contour would give some combination of $\Delta_1$ and $\Delta_2$.}

We notice that the preferred frame fields $u, \, s$ have not been used. But, if one wishes, they can be used. Indeed any couple of orthogonal fields $u,s$ which commute with $K_z$ and $K_t$ are related to $u^{\mathrm{ff}}, s^{\mathrm{ff}}$ by 
\begin{equation}
\label{gammaindSinapp}
u^{\mathrm{ff}} = (\Theta u + \gamma s )/H , \quad s^{\mathrm{ff}} = (\Theta s - \gamma  u)/H~,
\end{equation}
where the constant expansion is $\Theta= -  \nabla_\mu u^\mu$, and where the constant acceleration is $\gamma^\nu \doteq u^\mu  \nabla_\mu u^\nu = \gamma s^\nu$. 

\section{Nonseparability and Cauchy-Schwarz inequalities}
\label{app:CSclass}

In this appendix, we consider homogeneous Gaussian states. This implies that the state factorizes as 
\begin{equation}
\label{facto} 
\hat \rho = \bigotimes_{k > 0} \hat \rho_2^{(k)} ~, 
\end{equation}
where $\hat \rho_2^{(k)}$ fixes the state of the two-mode system $\bk, -\bk$. This also implies that $n_k$ and $c_k$ of \eq{eq:nkckdef} only depend on $\bk$. To be general, we work with $n_\bk \neq n_{-\bk}$, which means that the state is anisotropic. Our aim is to compare three inequalities relating the norm of $c_k$ to $n_\bk$ and $n_{-\bk}$ which allow to distinguish quantum from classical correlations, for a recent review, see e.g., Ref~\cite{Horodecki:2009zz}

\subsection{CS inequality in quantum mechanics}

Any quantum state (density matrix) $\hat \rho$ defines a (positive) scalar product on operators by:
\begin{equation}
(A,B)_\rho  \doteq \mathrm{Tr}\left (\hat \rho \hat A^\dagger \hat B    \right )~.
\label{scal}
\end{equation}
The corresponding Cauchy-Schwarz (CS) inequality implies
\begin{equation}
\label{eq:CSquantumst}
\abs{\mathrm{Tr}\left (\hat \rho \hat A^\dagger \hat B    \right )}^2\leq \mathrm{Tr}\left (\hat \rho\hat A^\dagger\hat A    \right ) \times \mathrm{Tr}\left (\hat \rho \hat B^\dagger \hat B    \right )~,
\end{equation}
When applied, to \eq{facto} with $\hat A= \hat a_\bk$ and $\hat B=\hat a_{-\bk}^\dagger$, one gets
\begin{equation}
\abs{c_k}^2 \leq n_\bk (n_{-\bk} +1)~.
\label{QCS}
\end{equation}
When $n_\bk = n_{-\bk}$, one obtains Eq.~\eqref{quantumCS}. 

\subsection{Separability} 

A bi-partite state is said separable~\cite{Werner:1989,Simon:2000zz} when it can be written as 
\begin{equation}
\label{eq:separablestate}
\hat \rho_{\rm sep}^{(k)} =  \sum_{n} p_n \,\hat \rho_{2, n}^{(k)}~,
\end{equation}
where $p_n\geq 0$, and where the two-mode states $\hat \rho_{2, n}^{(k)}$ are factorized $\hat \rho_{2, n}^{(k)} \doteq \hat \rho_n^{(\bk)} \otimes \hat \rho^{(-\bk)}_n$. The operators $\hat \rho_n^{(\pm \bk)}$ are density matrices for each one-mode system at fixed $\bk$. 

The structure of these states defines a new scalar product. It is given by
\begin{equation}
\label{eq:scalprodsep}
(X, Y)_{\rm sep} \doteq \sum_{n} p_n \, \mathrm{Tr}\left ( \hat \rho_{2, n}^{(k)} \hat X \right )^*  \mathrm{Tr}\left ( \hat \rho_{2, n}^{(k)} \hat Y \right )  ~, 
\end{equation}
where $\hat X$, $\hat Y$ are arbitrary operators. Considering operators that act on one sector only, i.e., $\tilde A = A \otimes 1$ and $\tilde B =1\otimes B$, one finds
\begin{subequations}
\label{eq:separableobs}
\begin{align}
\label{eq:scalofAB}
&\mathrm{Tr}\left (\hat \rho_{\rm sep} \tilde A^\dagger \tilde B    \right )  =  \sum_{n} p_n \, \overline{A}_n^{(\bk)\,^*}\  \overline{B}^{(-\bk)}_n = (\tilde A, \tilde B)_{\rm sep}~, \\
& \begin{array}{rl}
  \mathrm{Tr}\left (\hat \rho_{\rm sep} \tilde A^\dagger \tilde A    \right )& \hspace{-0.2cm} =  \sum_{n} p_n \,  \mathrm{Tr}\left (\hat \rho_n^{(\bk)} A^\dagger A    \right ) \\
  &\hspace{-0.2cm}\geq \sum_{n} p_n \abs{\overline{A}^{(\bk)}_{ n}}^2 = (\tilde A, \tilde A)_{\rm sep}~,
 \end{array}
\label{eq:normofA}
\end{align}
\end{subequations}
where the quantities with a bar are the expectation values involving only one-mode states 
\begin{equation}
\overline{C}^{(\bk)}_{n} \doteq \mathrm{Tr}\left ( \hat \rho_n^{(\bk)} \hat C \right ) ~.
\end{equation}
The inequality in Eq.~\eqref{eq:normofA} comes from the positivity of $\mathrm{Tr}(\hat \rho^{(\bk)}_n  \hat \xi^\dagger_n \hat \xi_n )$ applied to $\hat \xi_n=\hat A- \overline{A}^{(\bk)}_{ n} $, which gives
\begin{equation} 
\begin{split}
\mathrm{Tr}\left (\hat \rho_n^{(\bk)} A^\dagger A    \right ) &\geq \abs{\overline{A}^{(\bk)}_{ n}}^2 ~,  \\
\mathrm{Tr}\left (\hat \rho_n^{(\bk)} A A^\dagger    \right ) &\geq \abs{\overline{A}^{(\bk)}_{ n}}^2 ~.
\end{split}
\label{no-ord}
\end{equation}
The crucial point here is that the bound is insensitive to the ordering of $A$ and $A^\dagger$. Therefore, when applying the CS inequality associated with the scalar product of Eq.~\eqref{eq:scalprodsep}, i.e., $\abs{(X,Y)_{\rm sep}}^2\leq {(X,X)_{\rm sep}}\times {(Y,Y)_{\rm sep}} $, to $X = \hat a_\bk$ and $ Y =\hat a_{-\bk}^\dagger$, the strongest bound is
\begin{equation}
\label{eq:classCSineq}
\abs{c_k}^2 \leq n_\bk n_{-\bk}~.
\end{equation}
The only difference with \eq{QCS} is that $n_{-\bk}+1$ has been replaced by $n_{-\bk}$ by virtue of \eq{no-ord}. In conclusion, the inequalities of \eq{nonsep} characterize the quantum states which are nonseparable. 

\subsection{Subfluctuant mode}

We show that nonseparable states possess a subfluctuant mode whose variance is smaller than that of the vacuum. In the isotropic case, the proof can be found in Ref.~\cite{Campo:2008ju}. Below, we extend the proof to the anisotropic case $n_\bk \neq n_{-\bk}$.

To obtain the subfluctuant mode, we diagonalize the $2\times 2$ covariance matrix $\mathrm{Tr} ( \hat \rho _2^{(k)} \{W^\dagger, W\})$ with $W = \left ( a_{-\bk} ,a_\bk^\dagger \right )$ by a rotation, and not by a $U(1,1)$ transformation (a Bogoliubov transformation). The operators
\begin{equation}
\begin{split}
L_\bk &= \cos \xi \ep{-i\theta} a_{-\bk} +  \sin\xi \ep{i\theta} a_\bk^\dagger ~,  \\
S_\bk &= -\sin\xi \ep{-i\theta} a_{-\bk} +  \cos\xi \ep{i\theta} a_\bk^\dagger ~,
\end{split}
\end{equation}
define the super- and the subfluctuant mode, and the two angles are
\begin{subequations}
\begin{align}
\cos(2\xi) &= (n_\bk - n_{-\bk}) / \sqrt{(n_\bk - n_{-\bk})^2 + \abs{c_k}^2}~, \\
\theta &= \rm{arg}(c_k) /2~.
\end{align}
\end{subequations}

One verifies that $\mathrm{Tr}(\hat \rho \{ S_\bk, L_\bk^\dagger  \}) = 0$, and that the spread of the subfluctuant mode is 
\begin{equation}
\mathrm{Tr}(\hat \rho \{ S_\bk, S_\bk^\dagger  \}) =  n_\bk+ n_{-\bk} +1 - \sqrt{(n_\bk - n_{-\bk})^2 + 4 \abs{c_k}^2}~.
\end{equation}
Using Eq.~\eqref{eq:classCSineq}, one establishes that $\mathrm{Tr}(\hat \rho \{ S_\bk, S_\bk^\dagger  \}) < 1$
implies that the state is nonseparable. QED.

\section{Flux and long distance correlations} 
\label{app:BHdS}

The expressions for the asymptotic flux and the correlation pattern are both encoded in \eq{eq:GacomXX}. To obtain them, we need two things. Firstly, we need to characterize $G_{\rm ret}^\om$ from the asymptotic region down to the NHR. To this end, we should perform a WKB analysis of the stationary damped modes. Secondly, we need to connect the WKB modes with the high-momentum de Sitter--like physics taking place very close to the horizon. 

\subsection{WKB analysis}

At fixed $\om$, using Eq.~\eqref{eq:boxdiss}, $\Box_{\rm diss}\, \phi_{\rm dec} = 0$ implies that the decaying mode $\phi_{\rm dec}^\om$ obeys 
\begin{equation}
\begin{split}
\bigg [ &\left ( i \omega - \partial_X v  \right )\left ( i \omega -v \partial_X   \right ) + F^2(-\partial_X^2) \\
&-  \frac{\gamma(- \partial_X)  (i \omega - \sqrt{v} \partial_X \sqrt{v}) \gamma( \partial_X) }{\Lambda}   \bigg] \phi_{\rm dec}^\omega =0~.
\label{decmodeeq}
 \end{split}
\end{equation}
The mode $\phi_{\rm dec}^\om$ decays when displacing $X$ along the direction of the group velocity. Hence, on the right of the horizon, the outgoing $U$-mode decays when $X$ increases, while it decreases for decreasing $X< 0$ in the left region, see Fig~\ref{fig:caracter}. Hence, $U$-modes spatially decay on both sides when leaving the horizon.

As in the case of dispersive fields, we look for solutions of \eq{decmodeeq} of the form
\begin{equation}
\varphi(X) = \ep{ i \int^X dX' Q_\om(X') }~,
\end{equation}
where $Q_\om(X)$ is expanded in powers of the gradient of $v(X)$. To first order, \eq{decmodeeq} gives 
\begin{align}
(\omega - &v(X) Q_\om + i\Gamma)^2 - (F^2 - \Gamma^2) = \\
 &\nonumber -\frac{i}{2} \partial_X \partial_Q \left [  (\omega - v(X) Q_\om + i\Gamma)^2 - (F^2 - \Gamma^2)    \right ]~,
\end{align}
where the functions $\Gamma > 0$ of Eq.~\eqref{eq:Gammaasgamma} and $F$ are evaluated for $P = Q_\om$. The leading order solution, the complex momentum $Q^{(0)}_\om(X) \doteq P^C_\om(X)$, contains no gradient, and obeys the complex Hamilton-Jacobi equation
\begin{equation}
\label{eq:Q0}
\omega - v(X)  P + i\Gamma( P)=  \sqrt{F^2( P) - \Gamma^2( P)} \doteq \tilde F(P)~. 
\end{equation}
As expected, this equation gives Eq.~\eqref{eq:dispersion} since $\Omega = \om - v P$. To first order in the gradient, we get a total derivative 
\begin{align}
\label{eq:Q1}
Q^{(1)}_\om =  \frac{i}{2} \partial_X \log &\bigg [ \frac{\tilde F(P_\omega^C)}{\partial_\om P^C_\om} \bigg ]~. 
\end{align}
Combining Eq.~\eqref{eq:Q0} and Eq.~\eqref{eq:Q1}, we obtain the decaying WKB-mode
\begin{equation}
\varphi_{\rm dec}^\omega(X) = \frac{\ep{ -  \mathcal{I}_\om(X,X_0) } \times \ep{i \int^X_{X_0} dX' P_\omega(X')}   }{\sqrt{2 v^C_{\rm gr} \, \tilde F (P_\omega^C )  }}~.
\label{pW}
\end{equation}
To get this expression, we introduced $ v^C_{\rm gr}  = 1/\partial_\omega P^C_{\omega} $ which can be conceived as a complex group velocity. We also decomposed $P^C_\om$ into its real part $P_\om$, and its imaginary part $P^I_\om$. The oscillating exponential is the standard expression, while the decaying one is $\int dX P^I_\om$. The latter is equal to $\mathcal{I}_\om $ of Eq.~\eqref{eq:optdepthBH} when working to first order in $\Gamma/P$, which is here a legitimate approximation. A preliminary analysis, similar to Eq.~(A12) of Ref.~\cite{Coutant:2011in}, indicates that the corrections to \eq{pW} are bounded by $\mathcal{O} ( \frac{\omega^2}{\Lambda^2 \abs{1+v}^3} + \frac{g^2 \omega}{\Lambda (1+v)^2})$. Hence \eq{pW} gives an accurate description everywhere but in the central region \textit{III} defined by $\kappa X_{\rm trans}$ of \eq{Xtrans}. 

Using \eq{pW}, the $U$-mode contribution to the commutator is, for $\om > 0$, 
\begin{equation}		
\begin{split}
G_\mathrm{c}^\omega(X,X') &= \theta( \mathcal{I}_\om(X,X'))\, \varphi_{\rm dec}^{\omega}(X) \, 
\left (\varphi_{\rm grw}^{\omega}(X') \right )^* \\
& \quad + \theta( \mathcal{I}_\om(X',X))\, \varphi_{\rm grw}^{\omega}(X) \, \left (\varphi_{\rm dec}^{\omega}(X') \right )^*~,
\label{GcU}
\end{split}
\end{equation}
where the growing mode $\varphi_{\rm grw}^\omega$ satisfies \eq{decmodeeq} with the opposite sign for the last term which encodes dissipation. The expression for $\om < 0$ is given by $G_\mathrm{c}^{-\om} = - (G_\mathrm{c}^\omega)^*$ which follows from the imaginary character of $G_\mathrm{c}$ in $t,X$ space. We used the sign of $\mathcal{I}_\om $ in \eq{GcU} so that a similar expression is valid on the left of the horizon. Note also that \eq{GcU} cannot be used to estimate $G_\mathrm{c}^\om$ across the horizon because the WKB approximation fails in region \textit{III}. Note finally that Eq.~\eqref{GcU} is valid only for $\Lambda \abs{X-X'} \gg 1$.

Having characterized in quantitative terms the impact of dissipation, we now work in conditions such that the mode damping is negligible far away from this central region. That is, we work with $X, X'$ obeying 
\begin{equation} 
X_{\rm trans} \ll \vert X \vert \ll \sqrt{\Lambda/\kappa^3}~,
\label{niced}
\end{equation} 
where the upper limit comes from the neglect of the second term in Eq.~\eqref{eq:optdepthBHlimit}. Under these conditions, the anticommutator of \eq{eq:GacomXX} is, for $\om > 0$, given by 
\begin{equation}
\label{eq:Gacnc}
\begin{split}
G_{\rm ac}^\omega (X,X') =& (2n_\omega+1) \left [\varphi_{ R}^\omega (X)\, (\varphi_{ R}^\omega (X'))^* \right.\\
 &\quad \quad \quad \quad + \left.( \varphi_{ L}^{-\omega} (X))^* \, \varphi_{ L}^{-\omega} (X')\right ] \\
&+2 {\rm Re}\left [ c_\omega \, \varphi_{ R}^\omega (X) \, \varphi_{ L}^{-\omega}(X') \right ]~,
\end{split}
\end{equation}
where $n_\om$ and $c_\om$ are constant because we are far from region \textit{III}, and where the $R$ and $L$ \textit{out} modes live on one side of the horizon and have unit norm. Being undamped, they are either relativistic, or, more generally, dispersive WKB modes. In the former case, they thus behave in the regions of interest, namely $I_{R/L}$ and $\mathit{II}_{R/L}$, as
\begin{equation}
\begin{split}
\varphi_{ R}^\omega &\underset{\mathit{II} } \sim \theta(X) \frac{X^{i \omega/\kappa }  }{\sqrt{2\omega}} \underset{I} \sim   \theta(X) \frac{\ep{i \omega X /(1+v_R)}  }{\sqrt{2\omega /(1+v_R)}}  ~,  \\
\left (\varphi_{ L}^{-\omega} \right )^* &\underset{\mathit{II}} \sim  \theta(-X)  \frac{(-X)^{i \omega/\kappa }  }{\sqrt{2\omega}} \underset{I}\sim  \theta(-X) \frac{ \ep{- i \omega X /|1+v_{L}|} }{\sqrt{\vert 2\omega /(1+v_{L})\vert}}~, 
\end{split}
\label{retism}
\end{equation} 
where $v_{R(L)}$ is the asymptotic velocity in the region $R$ ($L$, where $1+v_L < 0$). As in de Sitter, the (positive unit norm) mode $ \varphi_{ L}^{-\omega}$ living in the $L$ region has a negative Killing frequency.

In \eq{eq:Gacnc}, $n_\om$ and $c_\om$ are unambiguously defined because the $R/L$ modes are normalized in regions $I_{R/L}$. Thus, they respectively define the spectrum emitted by the black hole, and the $\om$-contribution of the correlation across the horizon. To compute them, we should find the equivalent of \eq{eq:n_k,c_k}. To this end, we shall use \eq{eq:GacXXP}, and exploit the fact that their values are fixed in the domain of $P$ given in \eq{eq:dominP}. 

\subsection{Connection with de Sitter physics}
\label{sec:BHds:ds}

In \eq{eq:GacXXP}, we need (the $U$-mode contribution of) $G_{\rm ret}^\om(X,P_1)$ with $\abs{X} \gg  X_{\rm trans}$, since we are interested in the far away behavior of $G_{\rm ac}^\om$, and with $P_1 \gtrsim \sqrt{\kappa \Lambda}$, because the integrand vanishes for lower values of $P$. Since $P_\om(X) \ll P_1$, the retarded character
of \eq{eq:Gret-solution} is automatically implemented, which means that 
\begin{equation}
\label{eq:GretisGc}
G_{\rm ret}^\om(X,P_1)= (-i) \, G_\mathrm{c}^\om(X,P_1)~.
\end{equation} 
The commutator $G_\mathrm{c}^\om(X,P_1)$, on the one hand, obeys Eq.~\eqref{decmodeeq} in $X$, and on the other hand, behaves as in de~Sitter for $P_1 \gtrsim \sqrt{\kappa \Lambda}$, when $\om_{\rm max}^{\rm diss}$ of \eq{eq:ommdiss} obeys $\kappa/\om_{\rm max}^{\rm diss}\ll 1$. This second condition means that the high $P_1$ behavior is governed by Eq.~\eqref{eq:Gret-solution} and \eq{prepomega}.
 
For simplicity we consider the massless case of Eq.~\eqref{eq:numdispersion}, when $g^2 = 2$. In this model, in de Sitter, using the Unruh modes of Eq.~\eqref{eq:unruhmodedef}, the $U$-mode contribution is 
\begin{equation}
\begin{split}	
\label{eq:GcXPun}
G_{\rm c, \, \rm dS}^\omega(X,\bP) = \ep{- \mathcal{I}_0^P } \bigg[& \varphi_{U}^\omega (X) \, (\theta(\bP) \varphi^{\omega} (P))^* \\ 
 & - \left (\varphi_{U}^{-\omega} (X)\right )^* (\theta(-\bP)\varphi^{-\omega} (P))    \bigg ]~,
\end{split}
\end{equation}
where $\mathcal{I}_0^P$ is given in \eq{eq:opticaldepth}, and where we replaced its lower value $P_\om(X) \ll \sqrt{\kappa \Lambda}$ by $0$ because $X$ is taken sufficiently large. Using Eq.~\eqref{eq:UnruhRLchange}, we can reexpress Eq.~\eqref{eq:GcXPun} in the $R/L$ \textit{out} mode basis. For $\omega>0$ we get
\begin{equation}
\begin{split}
\label{eq:GcXPchi}
G_{\rm c, \, \rm dS}^\omega(X,\bP) = \ep{- \mathcal{I}_0^P } \bigg[& \chi_{R}^\omega (X) (\chi_{R}^{\omega} (\bP))^* \\
& - (\chi_{L}^{-\omega} (X))^* \chi_{L}^{-\omega} (\bP)    \bigg ]~.
\end{split}
\end{equation}
In this we recover that the commutator possesses the same expression if one uses the \textit{in} (Unruh) or the \textit{out} mode basis. 

Equation~\eqref{eq:GcXPchi} applies as such to the black hole metric in the regions $\mathit{II}_{R/L}$, $ \kappa X_{\rm trans} \ll \vert \kappa X \vert < D/2$, because $G_{\rm c, \, \rm BH}$ obeys the same equations, and its normalization is fixed by the equal time commutators. In fact, in these regions the normalized black hole modes $\varphi_R^\omega, 
\varphi_L^{-\omega} $ coincide with the modes $\chi_{R}^\omega, \chi_{L}^{-\omega}$ of \eq{RLm}. Then, the WKB character of $\varphi_R^\omega, \varphi_L^{-\omega} $ guarantees that Eq.~\eqref{eq:GcXPchi} applies further away from the horizon, in the regions defined by \eq{niced}. Hence, in these regions, we have 
\begin{equation}
\begin{split}
\label{eq:GcXPchiBH}
G_{\rm c, \,\rm BH}^\omega(X,\bP) =\ep{- \mathcal{I}_0^P } \bigg[& \varphi_{R}^\omega(X) (\chi_{R}^{\omega}(\bP))^* \\
& - (\varphi_{L}^{-\omega} (X))^* \chi_{L}^{-\omega} (\bP)    \bigg ]~.
\end{split}
\end{equation}
We kept the de~Sitter modes in $\bP$ space because only $|\bP| \gg {\kappa/\Lambda}$ contribute to \eq{eq:GacXXP}. Using Eq.~\eqref{eq:GretisGc}, inserting the above expression in Eq.~\eqref{eq:GacXXP}, and comparing the resulting expression with Eq.~\eqref{eq:Gacnc}, we get 
\begin{subequations}
\label{eq:ncac}
\begin{align}
(2n_\omega+1) =  \int d\bP_1 d\bP_2 & \,  \chi_{R}^{\omega \, *} (\bP_1) \chi_{R}^{\omega}(\bP_2)\\
\nonumber & \times e^{-\mathcal{I}_0^{P_1} -\mathcal{I}_0^{P_2} }  N^\omega(P_1,P_2) ~.  \\
2c_\omega = \int d\bP_1 d\bP_2 &\,  \chi_{R}^{\omega \, *} (\bP_1) \chi_{L}^{-\omega \, *}(\bP_2) \\
\nonumber &\times e^{-\mathcal{I}_0^{P_1} -\mathcal{I}_0^{P_2} }  N^\omega(P_1,P_2) ~.  
\end{align}
\end{subequations}
These expressions are identical to those evaluated in de Sitter. Hence, $n_\om$ and $c_\om$ are respectively given by Eqs.~\eqref{nom} and~\eqref{com}. Therefore, to leading order in $\kappa/\Lambda$, and for an environment at zero temperature, $n_\om$ and $c_\om$ retain their standard relativistic expressions. 

This means that the state of the outgoing modes when they leave the central region \textit{III}, and propagate freely, is the Unruh vacuum~\cite{Jacobson:1991gr,Jacobson:1993hn,Brout:1995rd}. This can be explicitly checked from \eq{eq:ncac} by reexpressing the \textit{out} modes $\chi_{R/L}^{\omega}$ in terms of the Unruh modes of \eq{eq:unruhmodedef}. In this case, one finds that the mean number of Unruh quanta $n_\om^{\rm Unruh}$ is given by, see Eq.~\eqref{eq:nucudef},
\begin{equation}
\begin{split}
(2n_\omega^{\rm Unruh}+1) = & \iint_0^\infty dP_1 dP_2   (\phi^{\omega } (P_1))^* \phi^{\omega}(P_2) \label{Unv} \\
&\quad \quad  \quad \times e^{-\mathcal{I}_0^{P_1} -\mathcal{I}_0^{P_2} }  N^\omega(P_1,P_2)\\ 
  = & \, 1 + O(\kappa/\Lambda) ~. 
\end{split}
\end{equation}
In other words, the role of the double integrals in \eq{eq:ncac} and \eq{Unv}, whose integrand explicitly depends on the actual ``trans-Planckian'' physics governed by $\Lambda$, $f(P)$, $\Gamma(P)$, is to implement the Unruh vacuum in dissipative theories. 

\bibliographystyle{../../biblio/h-physrev}
\bibliography{../../biblio/bibliopubli}

\end{document}